\documentclass[showpacs,aps,twocolumn,amsmath,amsfonts,floatfix,xcolor=dvipsnames]{revtex4-1}

\usepackage{graphicx}        
\usepackage{color}
\usepackage{amsmath}
\usepackage
	[colorlinks=true,
	urlcolor=blue,
	linkcolor=blue,
    citecolor=blue]
    {hyperref}
\usepackage{revsymb4-1}
\usepackage{rotating}
\usepackage{setspace}
\usepackage{multirow}
\usepackage{color}
\usepackage{xcolor}
\usepackage{wasysym}
\usepackage{bm}
\usepackage{fixmath}
\usepackage{amsbsy}
\usepackage{pifont}
\usepackage{placeins}
\usepackage{hhline}
\usepackage{tabularx}
\usepackage{ulem}
\usepackage{bbm}
\usepackage{colortbl}
\usepackage{ulem}
\begin{document}
 
\title{
	Hybrid-space density matrix renormalization group study \\
	of the doped two-dimensional Hubbard model}
 
\onecolumngrid
 
\author{G.\ Ehlers,$^1$ S.\ R.\ White,$^2$ and R.\ M.\ Noack$^1$} 
\affiliation{
    $^1$ Fachbereich Physik, 
    Philipps-Universit\"at Marburg, 
    35032 Marburg, Germany \\
    $^2$ Department of Physics and Astronomy, 
    University of California, Irvine, 
    California 92697, USA }

\date{\today}

\begin{abstract}

The performance of the density matrix renormalization group (DMRG) 
is strongly influenced by the choice of the local basis 
of the underlying physical lattice.
We demonstrate that, for the two-dimensional Hubbard model, 
the hybrid--real-momentum-space formulation
of the DMRG is computationally more efficient 
than the standard real-space formulation.
In particular, we show that the computational cost 
for fixed bond dimension of the hybrid-space
DMRG is approximately independent of the width of the lattice, in
contrast to the real-space DMRG, 
for which it is proportional to the width squared.
We apply the hybrid-space algorithm to calculate the ground state of the doped
two-dimensional Hubbard model on cylinders of width four and six sites;
at ${n=0.875}$ filling, the ground state exhibits a striped
charge-density distribution with a wavelength of eight sites
for both ${U/t=4.0}$ and ${U/t=8.0}$.
We find that the strength of the charge ordering
depends on ${U/t}$ and on the boundary conditions.
Furthermore, 
we investigate the magnetic ordering as well as the decay of the static 
spin, charge, and pair-field correlation functions.

\end{abstract}


\maketitle

\section{Introduction}
\label{sec:introduction}

Although the one-band Hubbard model~\cite{hubbard1963} in two
dimensions has long been touted as a
leading candidate for explaining the basic phenomenon of high-temperature
superconductivity in copper oxide planes~\cite{bednorz1986CuOHTC},
whether the unmodified model provides sufficient features
to do this is still controversial~\cite{imada2007HubbardHTC}.
In order to clear up this issue, enormous effort is being made to
obtain its phase diagram numerically~\cite{leblanc2015HubbardBenchmark}. 
As the model is doped away from half filling, rich behavior
emerges~\cite{chan2016HubbardPhaseDiagram}, with phases that include
antiferromagnetic ordering near half filling, a metallic phase for weak
on-site interaction, and a superconducting phase for moderate interaction.
Many details of the phase diagram, such as the presence of charge
and spin density stripes~\cite{white2003StripesHubbard,hager2005stripe},
which have been shown experimentally to play a role in
high-temperature superconducting materials
~\cite{tranquada1995StripedHTC,yoichi2002COStripes},
are yet to be unequivocally determined.

One of the most important numerical methods used to address these
questions is the  density matrix renormalization group
(DMRG)~\cite{white1992DMRG,schollwock2005dmrg,hallberg2006DMRG}.
In particular, the DMRG can provide unbiased results
that are a very useful benchmark and complement for other methods.
While the DMRG has been very successful in treating one-dimensional
systems and ladders, 
its application to two-dimensional systems, such as wider cylinders,
is much more challenging~\cite{stoudenmire2012dmrg2d}
because the growth of the entanglement between the DMRG blocks, which
is generally proportional to the system width for a short-range Hamiltonian,
leads to exponential growth of the computational cost.
Furthermore, the DMRG treats two-dimensional systems by mapping the
lattice to an intrinsically one-dimensional matrix product
state (MPS)~\cite{liang1994DMRG2d},
resulting in longer-range effective interaction even for short-range models.

More recently developed tensor network methods such as
multiscale entanglement renormalization 
(MERA)~\cite{vidal2008MERA,corboz2009MERA}
and projected entangled pair states
(PEPS)~\cite{verstraete2008PEPS,corboz2010PEPS}
avoid this restriction in principle by adapting the topology of the tensor
network to the entanglement structure of the system.
However, these methods suffer from the fact that the scaling of their
computational costs with the dimension of the Hilbert space treated is
much higher than that of the standard DMRG or other MPS-based algorithms.

Another ansatz to broaden the applicability of the DMRG for
two-dimensional systems is, instead of changing the topology of the
underlying network, to change the local basis of the physical
model~\cite{legeza2015TNSOrbitalOptimization}.
A change of basis can influence three characteristics 
that drastically influence the performance of the DMRG:
the entanglement between subsystems, the range and structure of the
interactions, and the number of quantities that are conserved within
the specific representation.

For instance, for translational invariant models, the momentum-space DMRG
takes advantage of the conserved momentum quantum number and
achieves a significant speedup for a fixed size of the truncated Hilbert
space~\cite{xiang1996momentumDMRG,nishimoto2002MomentumSpaceDMRG}.
However, the scaling of the size of the Hilbert space needed
to maintain a given fixed accuracy is problematic.
In particular,
the block entropy is zero in the noninteracting limit in the
momentum-space representation, as the noninteracting Fermi sea is a
product state.
Unfortunately, short-range interactions in real space become long-range
in momentum space so that all parts of the system 
become strongly entangled 
and the block entropy increases rapidly as the interaction is turned on; 
as a result, the entropy scales with the volume of the system
for all nonzero interaction strengths.
Thus, the computational cost scales exponentially in the {\em volume}
of the system, making treating systems of any significant size
prohibitively expensive~\cite{ehlers2015MomentumSpaceDMRG}.

Recently, it has been shown that, by choosing a mixed basis,
one can partially take advantage of the performance benefits 
of the momentum-space DMRG 
but also retain the beneficial entanglement scaling of the real-space
representation~\cite{motruk2016hybridDMRG}.
Our goal hence is to further explore this approach.
In particular, we apply the hybrid--real-momentum-space DMRG 
to the two-dimensional Hubbard model on a lattice with cylindrical topology.

The remainder of the paper is organized as follows:
in Sec.~\ref{sec:model_and_method}, we express the Hubbard model in
the hybrid-space representation,
discuss the structure of the matrix product operator (MPO),
and outline how real-space two-point correlation functions can be
calculated in hybrid space. 
In Sec.~\ref{sec:performance}, we discuss the computational cost of
the hybrid-space DMRG as applied to the two-dimensional Hubbard model.
In particular, we analyze the scaling of the CPU-time and memory costs
as a function of the cylinder width, 
and verify the results using realistic numerical calculations.
Section~\ref{sec:results} then describes our study of the ground
state of the doped Hubbard model  
at filling ${n=0.875}$ and interaction strengths ${U/t=4.0}$
and ${U/t=8.0}$ on cylindrical lattices of
width $4$ and $6$.
In particular, we address the questions of whether the ground state
exhibits stripe structures and whether the pairing correlations are enhanced.
Finally, Sec.~\ref{sec:conclusion} contains the conclusion.

\section{Model and method}
\label{sec:model_and_method}

\subsection{Hubbard model in hybrid space}
\label{sec:hubbard_model_in_hybrid_space}

We investigate the two-dimensional Hubbard model with 
nearest-neighbor hopping and on-site Coulomb repulsion
defined by the Hamiltonian
\begin{equation}
	H = 
	- t \sum_{\langle {\bf r} , {\bf r}' \rangle\,\sigma} 
	c^\dagger_{{\bf r}\,\sigma} \, c^{\vphantom{\dagger}}_{{\bf r}\,\sigma}
	+ U \sum_{\bf r} n^{\vphantom{\dagger}}_{{\bf r}\,\uparrow} \,
	n^{\vphantom{\dagger}}_{{\bf r}\,\downarrow}	
	\, ,
	\label{eqn:HubbardReal}
\end{equation}
where ${\langle {\bf r} , {\bf r}' \rangle}$ denotes nearest neighbors
on a square lattice with lattice sites ${{\bf r}=(x,y)}$.
Here, 
${c^{\vphantom{\dagger}}_{{\bf r}\,\sigma}}$ 
${(c^{{\dagger}}_{{\bf r}\,\sigma})}$
are  creation (annihilation) operators for electrons with spin
${\sigma \in \left\{\uparrow\,,\,\downarrow\right\}}$,
and
${n^{\vphantom{\dagger}}_{{\bf r}\,\sigma} =
  c^{\dagger}_{{\bf r}\,\sigma}\,c^{\vphantom{\dagger}}_{{\bf
      r}\,\sigma}}$ 
is the particle-number operator.
We take the lattice geometry to be cylindrical,
with cylinder length~$L_x$, width (i.e., circumference)~$L_y$,
a lattice spacing of unity,
and periodic or antiperiodic boundary conditions in the transverse direction.
Such a geometry is favorable for hybrid--real-momentum-space representation.
We Fourier transform in the transverse ($y$-) direction, i.e., write
\begin{equation}
	c^{{\dagger}}_{{\bf r}\,\sigma} = 
	\frac{1}{\sqrt{L_y}} \sum_{k_y} e^{- {\rm i} k_y  y } \;
	c^{{\dagger}}_{x\,k_y\,\sigma} 
	\label{eqn:Fourier}
\end{equation}
and correspondingly for ${c{^{\vphantom{\dagger}}_{{\bf r}\,\sigma}}}$.
For periodic boundary conditions, the transverse momentum points
are given by ${k_y= 2 \pi \, j/ L_y}$ with integer ${0 \leq j < L_y}$, 
whereas, for antiperiodic boundary conditions, 
${k_y= 2 \pi \, (j+\tfrac{1}{2})  / L_y}$. 
The resulting Hamiltonian in hybrid space,
\begin{subequations}
\begin{align}
	\label{eqn:HybbardTX}
	H  = & - t \sum_{\langle x,x' \rangle\,k_y\,\sigma} 
	c^\dagger_{x\,k_y\,\sigma} \, c^{\vphantom{\dagger}}_{x'\,k_y\,\sigma}\\
	\label{eqn:HybbardTY}
	 &  + \sum_{x\,k_y\,\sigma} \varepsilon(k_y) \;
	n^{\vphantom{\dagger}}_{x\,k_y\,\sigma} \\
	\label{eqn:HybbardU}
	  &  +
	\frac{U}{L_y} 
	\sum_{x\,k_y\,p_y\,q_y} 
	c^\dagger_{x\,k_y+q_y\,\downarrow} \, c^\dagger_{x\,p_y-q_y\,\uparrow} \,
	c^{\vphantom{\dagger}}_{x\,p_y\,\uparrow} \,
	c^{\vphantom{\dagger}}_{x\,k_y\,\downarrow} \, ,
\end{align}
\end{subequations}
consists of three terms,
a longitudinal hopping term~(\ref{eqn:HybbardTX}), 
a transverse hopping term~(\ref{eqn:HybbardTY})
with dispersion relation ${\varepsilon(k_y) = - 2 \, t \cos k_y}$,
and a long-range interaction term~(\ref{eqn:HybbardU}).
Note that the on-site Hubbard interaction becomes long-range in the
momentum direction, but remains short-range in the real-space direction.

As will be described in Sec.~\ref{sec:results},
we find states with striped charge and spin density patterns for
systems that are moderately doped away from half filling.
In order to stabilize and target states with a particular 
wavelength of the charge-density stripes, $\lambda$,
we sometimes apply an additional pinning field
\begin{equation}
	H_n =
	P \sum_{x\,k_y\,\sigma} 
	\sin(\phi + \kappa x) \;
	n^{\vphantom{\dagger}}_{x\,k_y\,\sigma}
	\, 
	\label{eqn:PinningField}
\end{equation}
to the doped system.
The field couples directly to the local charge density in hybrid space
and can be tailored using its amplitude $P$, 
wavenumber ${\kappa=2\pi\,/\,\lambda}$, and
phase $\phi$.
Depending on the stability of the target state,
we either turn of the pinning field after the initial sweeps of the DMRG,
or we keep the field amplitude finite but small throughout the calculation
and subtract the contribution of the field to the ground-state energy 
once the calculation is completed.
In order to ascertain that the presence of a small pinning field has
no significant effect on the physical results, we have compared 
calculations with and without a pinning field 
for cases in which the pinning field can be ramped to zero during the
calculation without affecting the final convergence
and have found the difference in observables and in energy (with the
energy of the pinning field subtracted) to be negligible.

\subsection{Hybrid-space matrix product operator}
\label{sec:hybrid-space_mpo}

In its modern formulation, the DMRG algorithm is best described within
the framework of MPSs and MPOs~\cite{rommer1995MPS,schollwock2011MPS}.  
The MPS and MPO store the coefficients of the state
${\left|\Psi\right>}$ and Hamiltonian $H$ as products of matrices
associated with each site of the DMRG chain.
The MPS and MPO bonds are the contractions of the row and column
indices, respectively, of
neighboring matrices.
The dimensions of the MPS bonds, $m$,
are directly related to the number of
states kept in the traditional DMRG,
and the dimensions of the MPO bonds refer to the operators stored in
the left and right block of the DMRG.
The MPO also encodes the rules of how ${H\left|\Psi\right>}$ must be calculated 
within the DMRG-specific block-site-site-block decomposition of the system
and how the DMRG blocks are updated during the sweeping process.
Therefore, it is crucial to find the optimal MPO representation for a
given model and lattice, i.e., to find the MPO with minimal bond dimension.

Roughly speaking, the dimension of each bond of the MPO depends on the
number of terms of the Hamiltonian acting simultaneously on both sites
of the bond.
If the Hamiltonian is factorizable in an appropriate way,
the bond dimension can be minimized by accumulating all interactions
between opposing sides of a ``composed'' operator,
so that the same interaction can be expressed using fewer 
terms~\cite{xiang1996momentumDMRG}.
In this section, we describe how this is done in principle;
the details of the construction of the MPO matrices for the hybrid-space 
Hubbard Hamiltonian are given in
Appendix~\ref{app:mpo}.

For the two-dimensional Hubbard model in real space,
the bond dimension of the optimal MPO is proportional to the system
width ${L_y}$.
This can be understood as follows:
\begin{figure}[htpb]
	\includegraphics[width=6.0cm]{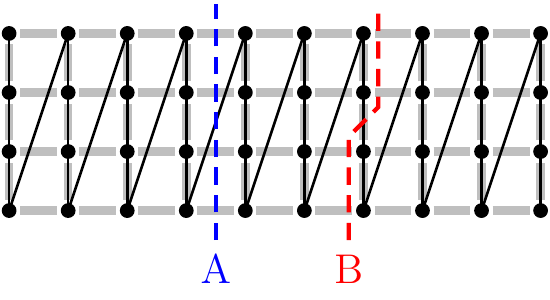}
	\caption{
		(Color online) 
		Mapping of the one-dimensional DMRG chain (black solid line) 
		onto a ${10{\times}4}$ square lattice.
		The bold gray lines depict nearest neighbors on the square lattice, 
		and the dashed lines A and B depict two possible cuts 
		through the DMRG chain
		and the corresponding bipartitions of the system.
	}
	\label{fig:mapping}
\end{figure}
Figure~\ref{fig:mapping} shows the most common mapping 
of a two-dimensional square lattice 
onto the one-dimensional MPS / MPO chain; 
the one-dimensional path is simply folded over the width of the
lattice into two dimensions.
Therefore, it is clear that, whenever the chain is cut, 
the number of nearest-neighbor bonds, and thus the number of hopping
term in the two-dimensional Hubbard model in real space,  is proportional to the
width of the system.
Since (almost) all of these terms act on different sites of the lattice, 
the MPO must include an individual channel for each term;
thus its bond dimension is proportional to the system width.

In hybrid space, the situation is more complicated.
In particular, the bond dimension of the MPO depends on where we
cut the system:
if the cut is between two neighboring rings of the cylinder 
(dashed line A in Fig.~\ref{fig:mapping}),
only real-space-like hopping terms~(\ref{eqn:HybbardTX}) are cut,
and the bond dimension is again proportional to the system width.
If the cut separates sites of the same ring of
the cylinder (dashed line B in Fig.~\ref{fig:mapping}),
the corresponding MPO bond has to carry all the long-range
interactions in the term~(\ref{eqn:HybbardU});
for each ring this sum runs over three independent momenta.
If we were to construct the MPO with individual channels for each
term, the resulting bond dimension would be proportional to
${L_y^3}$.

Fortunately, for a given bipartition,
the sum~(\ref{eqn:HybbardU}) can be factorized
to reduce the effective number of terms as follows:
for fixed $x$, the part of~(\ref{eqn:HybbardU}) for which the
annihilation and creation operators 
are in separated parts of the system can be rewritten as
\begin{equation}
	\frac{U}{N} \sum_{k_y} 
	A^\dagger_{x\,k_y} \,
	A^{\vphantom{\dagger}}_{x\,k_y}	
	\label{eqn:HubbardHybrid}
\end{equation}
with composed operators
\begin{align}
	A^\dagger_{x\,k_y} & = \sum_{p_y}
	c^\dagger_{x\,p_y\,\downarrow} \, 
	c^\dagger_{x\,k_y-p_y\,\uparrow} \, , \nonumber \\	
	A^{\vphantom{\dagger}}_{x\,k_y} & = \sum_{p_y}
	c^{\vphantom{\dagger}}_{x\,p_y\,\uparrow} \,
	c^{\vphantom{\dagger}}_{x\,k_y-p_y\,\downarrow} \, .
\end{align}
Analogous steps can be carried out for all possible distributions of
creation and annihilation operators of both spin species onto the two
subsystems, resulting in a formulation in which the total number of terms
connecting the two parts of the system is ${\mathcal{O}(L_y)}$.
Thus, without approximation, we obtain an MPO in which the dimension
of each bond is proportional to the system width, just as in real space.

This approach was introduced for the momentum-space DMRG 
in Ref.~\cite{xiang1996momentumDMRG}.
The technique can be applied to other models as long as the
Hamiltonian is factorizable.
The hybrid-space Hamiltonian and the 
optimized MPO for the fermionic
Hofstadter model are described in detail in 
Ref.~\cite{motruk2016hybridDMRG}.
Similar optimizations of MPOs have also been carried out 
for other systems with long-range interactions, 
such as the quantum chemical Hamiltonian~\cite{keller2015}.

The exact bond dimension for the Hubbard model in hybrid space varies
with the position of the cut in the cylinder, whereas in real space it is
constant within the bulk of the system.
\begin{table}[htpb]
	\def\arraystretch{1.25}
	\centering
	\caption{ 
		Average MPO bond dimension for the two-dimensional Hubbard model 
		in the real-space and in the hybrid-space representation
		for different cylinder widths.		
	}
	\begin{tabularx}{8.6cm}{
		>{\centering\arraybackslash}p{2.1cm} 
		>{\centering\arraybackslash}p{1.45cm} 
		>{\centering\arraybackslash}p{1.45cm}
		>{\centering\arraybackslash}p{1.45cm} 
		>{\centering\arraybackslash}p{1.45cm}}
		\hline\hline & $L_y = 4$ & $L_y = 6$ & $L_y = 8$  & $L_y = 10$ \\
		\hline Real space & $18$ & $26$ & $34$ & $42$ \\
		Hybrid space & $26.0$ & $45.7$ & $66.5$ & $84.6$ \\
		\hline\hline
	\end{tabularx}
	\label{tab:mpo_dimensions}
\end{table}
Assuming a constant MPS bond dimension,
a good indicator for the computational cost of the DMRG 
is the average MPO bond dimension,
which is approximately twice as large in the hybrid-space representation
as in the real-space case (Table~\ref{tab:mpo_dimensions}).
Note that the averaged MPO bond dimension for the hybrid-space MPO may
still vary slightly  
for different orderings of the momentum points of each cylinder ring
within the DMRG chain.

\subsection{Real-space correlation functions in hybrid space}
\label{sec:correlation_functions_in_hybrid_space}

The equal-time
spin, charge, and pair-field correlation functions are defined as
\begin{align}
	S({\bf r},{\bf r}') & = 
	\langle S_{\rm z}({\bf r}) \, S_{\rm z}({\bf r}') \rangle \, , \nonumber \\
	C({\bf r},{\bf r}') & = 
	\langle n({\bf r}) \,  n({\bf r}') \rangle 
	- \langle n({\bf r}) \rangle \, 
	\langle   n({\bf r}') \rangle \, , \nonumber \\
	D_{\rm y\,y}({\bf r},{\bf r'}) & = 
	\langle \Delta^{\dagger}_{\rm y} ({\bf r}) \,
	\Delta^{\vphantom{\dagger}}_{\rm y} ({\bf r'}) \rangle
	\label{eqn:correlationfunctions}
\end{align}
with
\begin{align}
	S_{\rm z}(x,y) & =  
	 c^{\dagger}_{x\,y\,\uparrow} \, 
	 c^{\vphantom{\dagger}}_{x\,y\,\uparrow}
	-c^{\dagger}_{x\,y\,\downarrow} \,
	 c^{\vphantom{\dagger}}_{x\,y\,\downarrow} \, , \nonumber \\
	n(x,y) & =  
	 c^{\dagger}_{x\,y\,\uparrow} \, 
	 c^{\vphantom{\dagger}}_{x\,y\,\uparrow}
	+c^{\dagger}_{x\,y\,\downarrow} \,  
	 c^{\vphantom{\dagger}}_{x\,y\,\downarrow} \, , \nonumber \\
	\Delta^{\dagger}_{\rm y} (x,y) & =  
	\frac{1}{\sqrt{2}} \, 
	(c^{\dagger}_{x\,y+1\,\uparrow} \, c^{\dagger}_{x\,y\,\downarrow}
	-c^{\dagger}_{x\,y+1\,\downarrow} \, c^{\dagger}_{x\,y\,\uparrow}) \, ,
\end{align}
where ${S_{\rm z}(x,y)}$ measures the local spin, 
${n(x,y)}$ is the local charge density,
and ${\Delta^{\dagger}_{\rm y} (x,y)}$ is the creation operator for a
pair of spin-up and -down particles on sites ${(x,y)}$ and ${(x,y+1)}$. 

Measuring two-point real-space correlation functions in hybrid space
raises the same issues as implementing the MPO of the Hamiltonian:
applying the Fourier transformation~(\ref{eqn:Fourier}) to the
correlation functions~(\ref{eqn:correlationfunctions}) 
introduces sums over multiple momenta similar to term~(\ref{eqn:HybbardU}).
These sums can again be factorized analogously to Eq.~(\ref{eqn:HubbardHybrid}):
the pair-field correlation functions in hybrid space can be written as
\begin{equation}
	D_{\rm y\,y}({\bf r}, {\bf r'}) = 
	 \frac{1}{2\,L_y^2}\sum_{k_y} \, e^{{\rm i} k_y (y-y')}  \;
	 \langle O_{\rm y}^{\dagger}(x,k_y) \, 
	 O_{\rm y}^{\vphantom{\dagger}}(x',k_y) \rangle
\end{equation}
with composed operators
\begin{align}
  O_{\rm y}^{\dagger}(x,k_y) =
  2\,\sum_{p_y} \cos(p_y-k_y/2) \;  
  c^{\dagger}_{x\,p_y\,\uparrow} \,
  c^{\dagger}_{x\,k_y-p_y\,\downarrow}  \nonumber \, , \\ 
    O_{\rm y}^{\phantom{\dagger}}(x,k_y) =
    2\,\sum_{p_y} \cos(p_y-k_y/2) \;  
    c^{\phantom{\dagger}}_{x\,p_y\,\uparrow} \,
    c^{\phantom{\dagger}}_{x\,k_y-p_y\,\downarrow} \, . 
\end{align}
The spin and charge correlation functions can be treated analogously.
Thus, the correlation functions~(\ref{eqn:correlationfunctions})
can be measured without changing the scaling of the total
computational cost of the algorithm.

\section{Performance}
\label{sec:performance}

Here we start by emphasizing that the dimension of
the MPS bonds required to obtain a given accuracy scales
exponentially with the block entropy; thus, the scaling of the block
entropy with system parameters and lattice size has a decisive
influence on the performance for fixed accuracy.
For two-dimensional systems
to which the entropy area law~\cite{eisert2010areaLaw}
applies, the entropy is proportional to system width $L_y$.
The resulting exponential scaling of the computational cost 
with system width
is a fundamental limitation of the hybrid-space DMRG as well as of the
real-space DMRG and all known MPS-based algorithms.

In this section, we at first neglect the variation in the required bond
dimension $m$ on model parameters and lattice size and analyze the
scaling of the computational
cost of the  real-space and hybrid-space DMRG for fixed $m$.
We thus estimate the performance gain of the hybrid-space
relative to the real-space representation.
We then compare our estimate to measurements of the actual
runtime and memory usage for typical calculations
and finally come back to the issue of the dependence of
the relative accuracy of the hybrid-space and real-space
representations on bond dimension $m$.

\subsection{Estimated scaling of the computational cost}
\label{sec:estimated_performance}

The majority of the computational cost in the DMRG algorithm comes
from applying the Hamiltonian to a state, ${H|\Psi\rangle}$; this is
the fundamental step in iterative eigensolvers such as the Lanczos or
Davidson algorithms~\cite{lanczos1950,davidson1975}.
Therefore, we examine the scaling of the operations required for this operation
with the dimension of the physical lattice sites, $d$,
cylinder width $L_y$, cylinder length $L_x$, 
and MPS bond dimension $m$ based on the structures of the MPS and MPO used.
The computational costs of other operations, e.g., 
changing the active sites of the DMRG in the sweeping process,
benefit from the hybrid-space representation in the same way.

We assume that the bond dimension of the MPO is proportional to $L_y$
in both real-space and hybrid-space representations, 
as is the case for the Hubbard model (see Sec.~\ref{sec:hybrid-space_mpo}).
In our estimate, we neglect the possibility of exploiting the
conservation of spin and charge quantum numbers, as it would have the
exact same effect on the computational costs in both representations.

In the real-space representation,
one Lanczos step then requires ${\mathcal{O}(d^2 \, L_y)}$ 
multiplications of ${m{\times}m}$ matrices,
resulting in ${\mathcal{O}(d^2 \, L_y \, m^3)}$ operations per Lanczos step
and ${\mathcal{O}(d^2 \, L_y^2 \, L_x \, m^3 \, K)}$ operations per DMRG sweep, 
with a fixed number of Lanczos steps per DMRG step, $K$.
The corresponding memory costs are ${\mathcal{O}(d^2 \, m^2 \, K)}$ 
for the $K$-dimensional Krylov space and 
${\mathcal{O}( L_y \, m^2 )}$ for the left and right DMRG block 
in the current block-site-site-block configuration.

Next, we analyze the scaling of the costs in the hybrid-space
representation under the assumption that the hybrid-space MPS has the
same total bond dimension $m$.
Since the Hamiltonian (\ref{eqn:HubbardHybrid}) conserves the
transverse momentum quantum number,
every matrix of the MPS can be written in a block-diagonal form
with $L_y$ blocks of size ${m'{\times}m'}$ with ${m'\approx m/L_y}$,
where every block corresponds to one momentum sector.
Therefore, the computational costs reduce to 
${\mathcal{O}(d^2 \, L_y^{-1} \, m^3)}$ per Lanczos step
and ${\mathcal{O}(d^2 \, L_x \, m^3 \, K)}$ per DMRG sweep,
and the memory costs become ${\mathcal{O}(d^2 \, L_y^{-1} \, m^2 \, K)}$ 
for the Krylov space
and ${\mathcal{O}(m^2)}$ for the DMRG blocks.
A side-by-side comparison of all costs 
is given in Table~\ref{tab:computational_costs}.

Note that the above argument is only valid 
because the MPS bonds decompose into $L_y$ momentum quantum number
sectors of approximately equal size ${m'}$.
For other conserved quantities such as the
total spin or particle number,
there typically is  a different distribution of the quantum number sectors,
with only a few large sectors dominating the computational costs of
the algorithm; in these cases, the speedup depends primarily
on the size of the largest sectors rather than on the number of sectors.

\begin{table}[htpb]
	\def\arraystretch{1.25}
	\caption{ 
		Scaling of the runtimes of a single Lanczos step,
		${T_{\text{Lanczos}}}$, 
		and of a DMRG sweep, ${T_{\text{sweep}}}$,
		and scaling of the memory costs associated with the Krylov space,
		${M_{\text{Krylov}}}$, 
		and the left and right DMRG blocks of the current 
		block-site-site-block configuration, ${M_{\text{block}}}$,
		as a function of the cylinder length $L_x$, the cylinder width $L_y$,
	    the MPS bond dimension $m$, 
		the local lattice dimension $d$, and the dimension 
		of the Krylov subspace, $K$.
	}
	\begin{tabularx}{8.6cm}{
		>{\centering\arraybackslash}p{2cm} 
		>{\centering\arraybackslash}p{3cm}  
		>{\centering\arraybackslash}p{3cm}}
		\hline\hline & Real space & Hybrid space \\ 
		\hline $T_{\text{Lanczos}}$ & 
		$\mathcal{O}(d^2 \, L_y \, m^3)$ & 
		$\mathcal{O}(d^2 \, L_y^{-1} \, m^3)$ \\
		$T_{\text{sweep}}$ 	& 
		$\mathcal{O}(d^2 \, L_y^2 \, L_x \, m^3 \, K)$ & 
		$\mathcal{O}(d^2 \, L_x \, m^3 \, K)$  \\
		$M_{\text{Krylov}}$ &  
		$\mathcal{O}(d^2 \, m^2 \, K)$ & 
		$\mathcal{O}(d^2 \, L_y^{-1} \, m^2 \, K)$  \\
		$M_{\text{block}}$ & 
		$\mathcal{O}( L_y \, m^2 )$ & 
		$\mathcal{O}( m^2 )$  \\
		\hline \hline
	\end{tabularx}
	\label{tab:computational_costs}
\end{table}

In conclusion, the computational costs of the hybrid-space DMRG are expected 
to be independent of $L_y$ for fixed $m$, 
whereas for the standard real-space DMRG
the runtime and the memory consumption scales as ${L_y^2}$ and $L_y$,
respectively (Table~\ref{tab:computational_costs}).
Estimating the bulk MPO bond dimension for the Hubbard model to be
approximately twice as large in hybrid space as in real space
(Table~\ref{tab:mpo_dimensions}),  
we estimate a total speedup of roughly ${L_y^2 / 2}$.

\subsection{Measured performance}
\label{sec:measured_performance}

In order to investigate the actual performance,
we compare the computational costs of real-space and hybrid-space
calculations for Hubbard cylinders with length $16$, widths $4$,
$6$, and $8$, and periodic transverse boundary conditions.
All calculations were carried out using the same code 
and using six physical cores on an
Intel\textsuperscript{\textregistered}
Xeon\textsuperscript{\textregistered} X5650 CPU.
In both representations, we exploit the block-diagonal structure of
the MPS and MPO matrices with respect to the conserved charge and
total spin quantum numbers;
in the hybrid space version, we further decompose the matrices using
the transverse momentum quantum number.

\begin{figure}[htpb]
	\includegraphics[width=8.6cm]{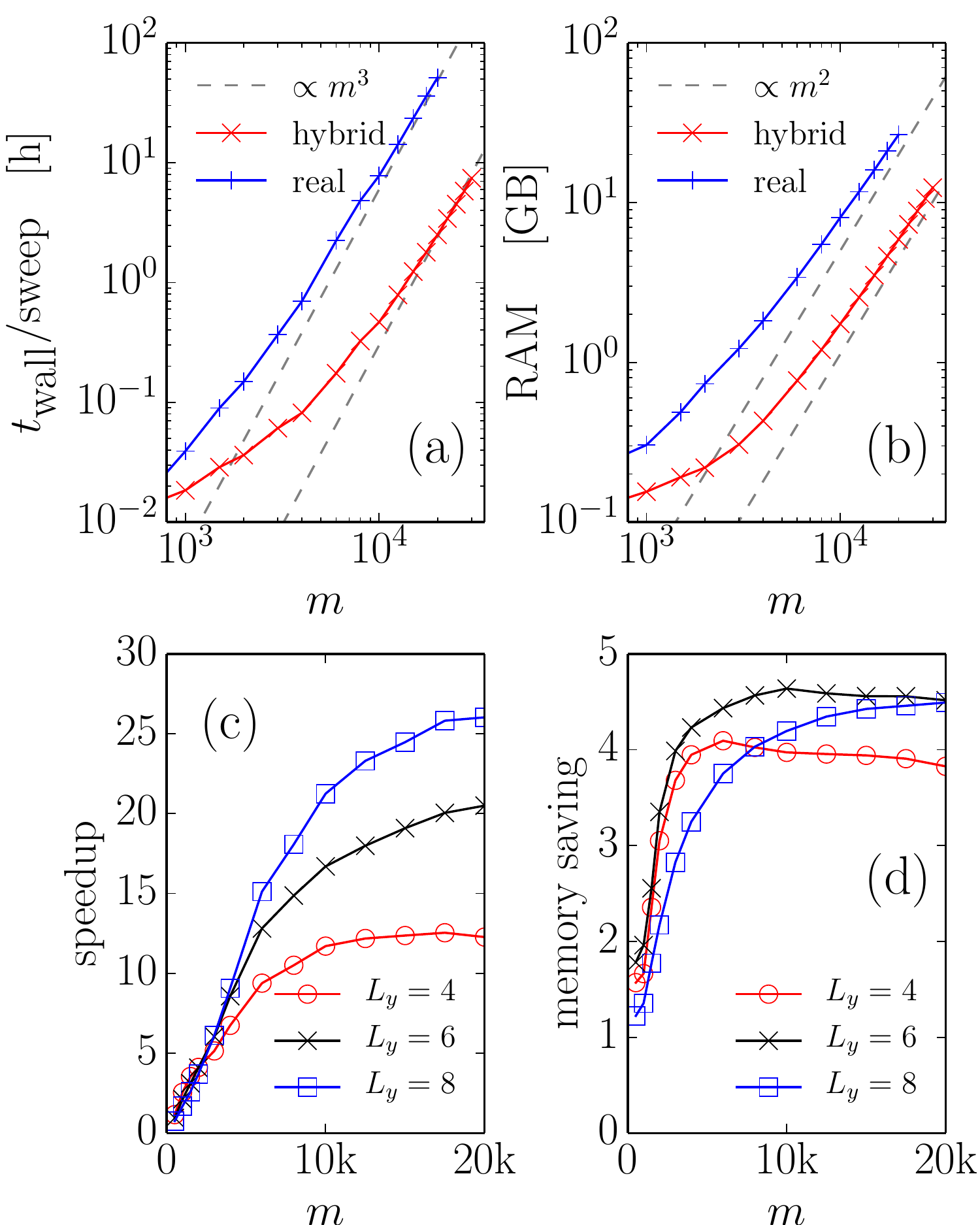}
	\caption{
		(Color online) Performance comparison between real-space 
		and hybrid-space DMRG, calculated for ${U/t=4.0}$ and ${n=0.875}$;
		(a) wall time per DMRG sweep and (b) peak memory consumption
		for a ${16{\times}6}$ cylinder as a function 
		of the MPS bond dimension $m$.
		The dashed gray lines depict the $m^3$ and $m^2$ scaling 
		expected in the ${m\rightarrow\infty}$ limit.
		(c) Speedup and (d) memory savings
		of the hybrid-space DMRG compared to real-space DMRG
		as a function of $m$ for ${16{\times}4}$, 
		$16{{\times}6}$, and ${16{\times}8}$
                cylinders. 
	}
	\label{fig:performance_comparison}
\end{figure}
Figure~\ref{fig:performance_comparison} shows a comparison of the runtime 
and memory requirements
for different $L_y$ as a function of $m$.
As expected, the CPU time per sweep is proportional 
to $m^3$ for large $m$ in both cases, 
while the peak memory consumption scales with
$m^2$~[Figs.~\ref{fig:performance_comparison}(a)~and~\ref{fig:performance_comparison}(b)].
The deviation from this limiting behavior for smaller $m$ is due to
the quantum number bookkeeping and other overhead in the code and is
amplified for the hybrid-space DMRG because of the additional momentum
quantum number.
In agreement with the predictions above, the speedup of the
hybrid-space DMRG over the real-space calculations
is larger for wider cylinders [Fig.~\ref{fig:performance_comparison}(c)]. 
Because of the additional overhead, the full speedup of the
hybrid-space code is only seen at large $m$.
In the regime in which both methods still provide results within
reasonable time in our calculations, the hybrid-space DMRG is
approximately $12$ times faster for ${L_y=4}$ and up to $20$ and $26$
times faster for ${L_y=6}$ and ${L_y=8}$.
In terms of the peak memory consumption, we observe a weaker influence of $L_y$:
the memory savings for the larger $m$ values treated varies only
between a factor of $4$ for ${L_y=4}$ and $4.5$ for ${L_y=6}$ and
${L_y=8}$ [Fig.~\ref{fig:performance_comparison}(d)].  

\begin{figure}[htpb]
	\vspace{0.25cm}
	\includegraphics[width=8.6cm]{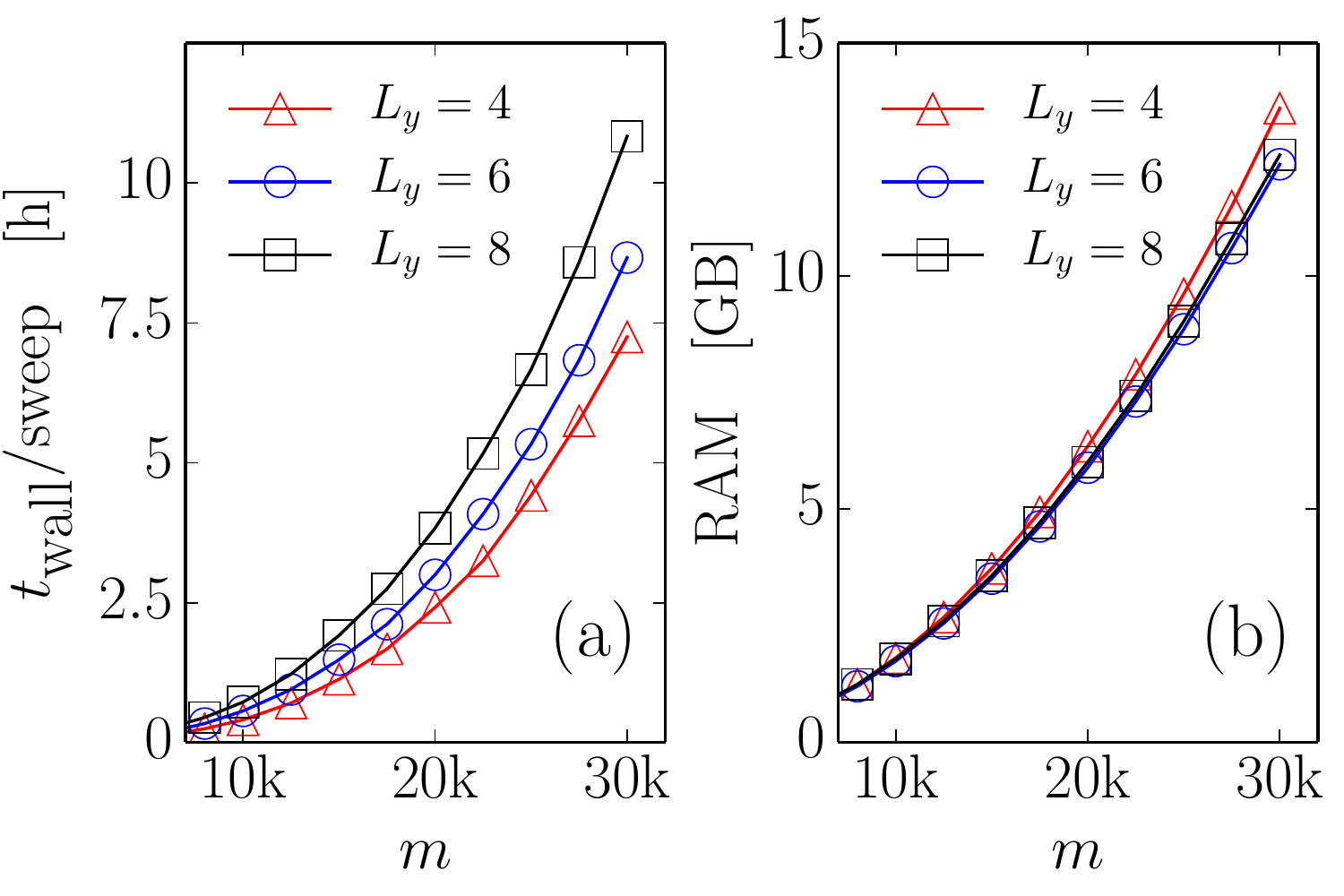}
	\vspace{-0.25cm}
	\caption{
		(Color online)
		(a) CPU time per sweep and (b) peak memory consumption 
	    of hybrid-space DMRG calculations
		for Hubbard cylinders with length 16 and width $L_y$
		as a function of the MPS bond dimension $m$.
		The calculations were carried out 
		for ${n=0.875}$ and ${U/t=4.0}$.
	}
	\label{fig:circumference_scaling}
\end{figure}
In Fig.~\ref{fig:circumference_scaling}, we compare the CPU time per sweep 
and the memory consumption of the hybrid-space DMRG for different $L_y$.
In agreement with Table~\ref{tab:computational_costs},
the computational costs are almost independent of $L_y$.
The minor growth of the runtime that is still present 
is clearly sublinear in $L_y$.
The measured computational costs deviate somewhat 
from the estimated costs even in the large-$m$ limit,  
in which the influence of overhead should be insignificant.
In particular,
the measured absolute runtime of the hybrid-space DMRG 
is not strictly independent of $L_y$, 
and thus the observed speedup over real-space DMRG grows 
more slowly than $L_y^2$.
This deviation is caused by fact that the MPO dimension 
is not perfectly proportional to $L_y$ for small $L_y$ 
(Table~\ref{tab:mpo_dimensions}).

\begin{figure}[htpb]
	\includegraphics[width=8.6cm]{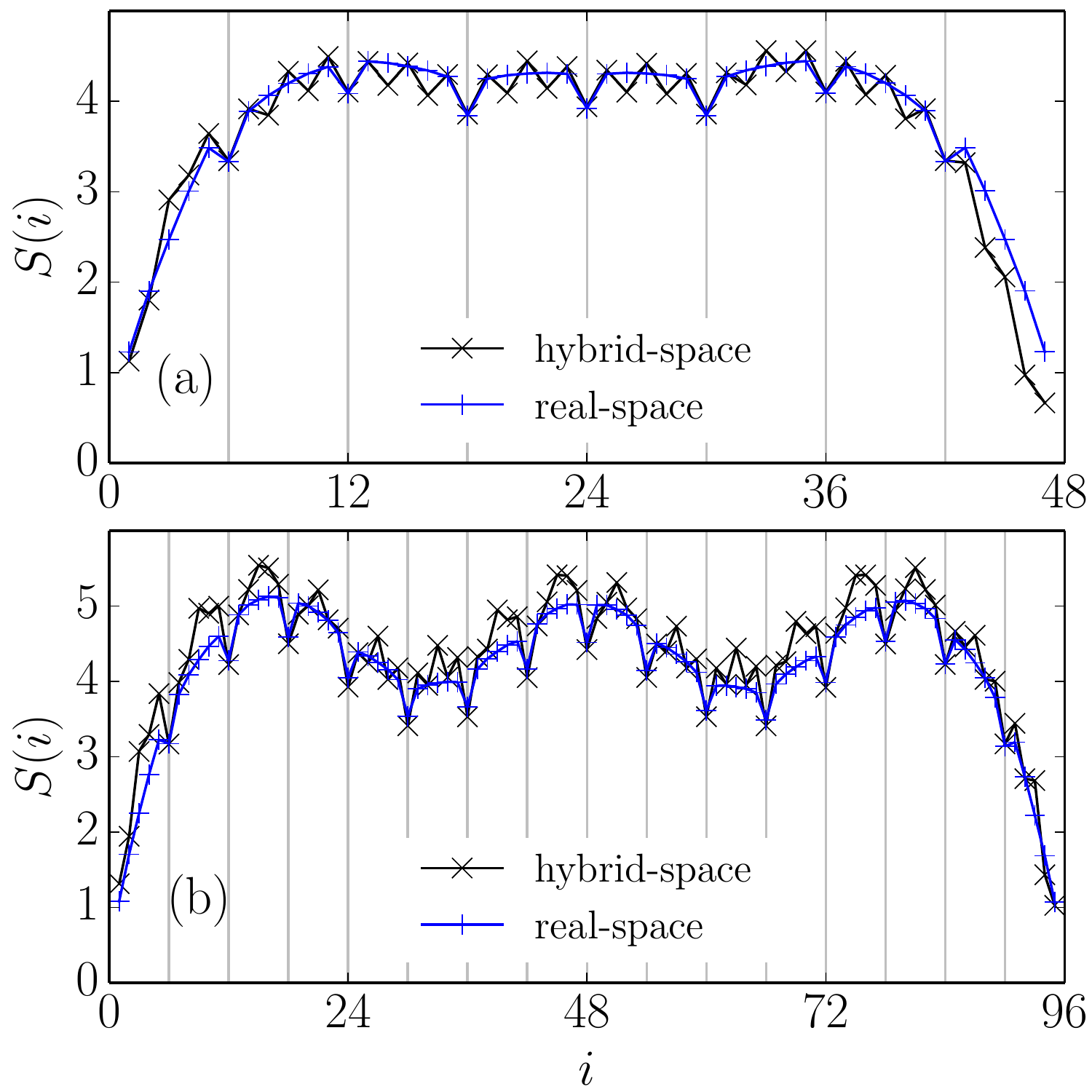}
	\caption{
	    (Color online) Von Neumann entropy ${S(i)}$ 
	    as a function of the MPS bond index $i$ for
   		(a) ${8{\times}6}$ cylinders at ${U/t=4.0}$ and	half-filling and
   		(b) ${16{\times}6}$ cylinders at ${U/t=8.0}$ and ${n=0.875}$.
	    The sites of the real-space and hybrid-space lattices 
	    are mapped to the MPS sites
	    in an $x$-direction-first ordering;
	    accordingly, the gray vertical lines indicate cuts between
        neighboring rings of the cylinders.
	}
	\label{fig:blockentropy}
\end{figure}

\begin{figure}[htpb]
	\includegraphics[width=8.6cm]{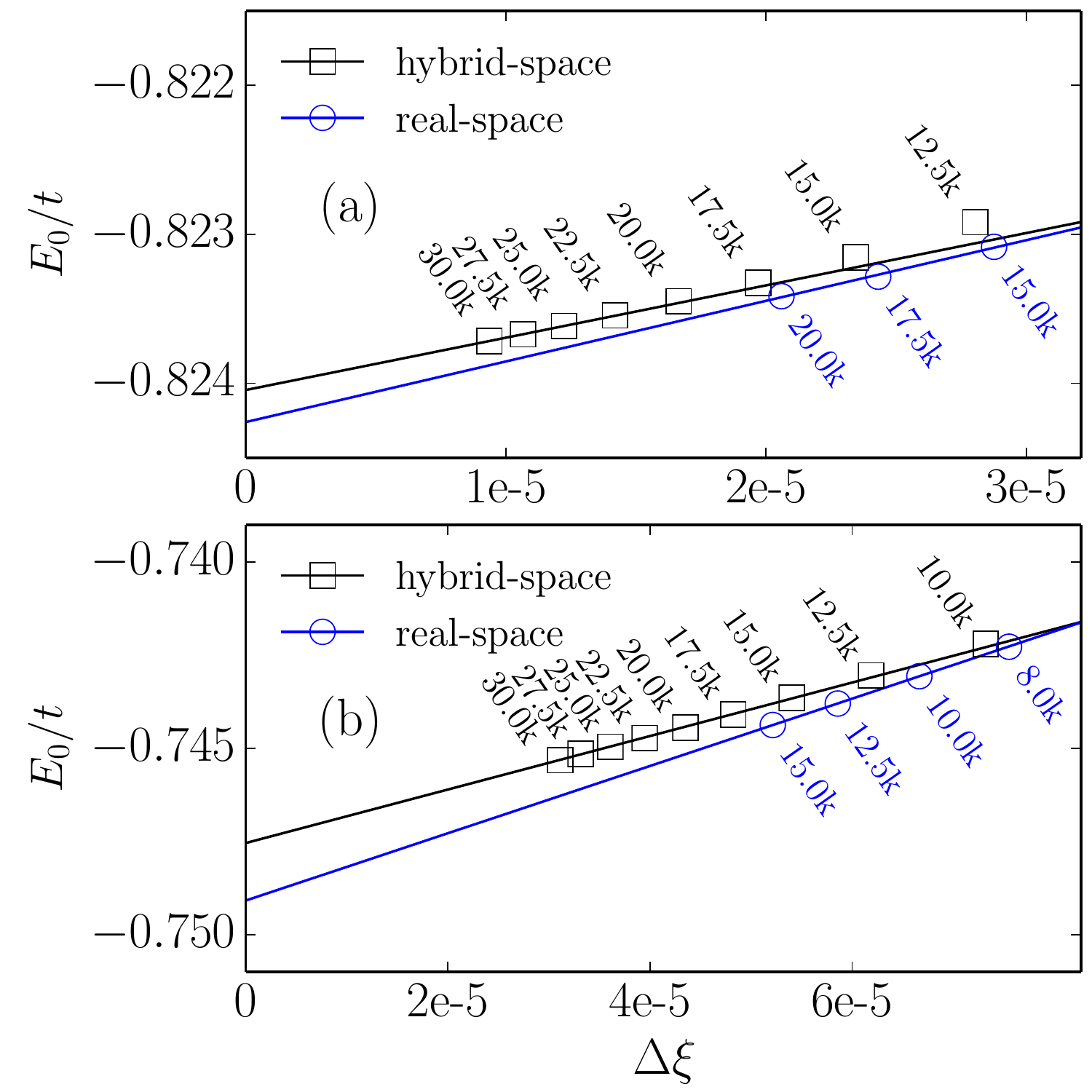}
	\caption{
		(Color online) Ground-state energy 
		obtained from hybrid-space DMRG and real-space DMRG
		as a function of the discarded weight per site, ${\Delta\xi}$, for 
		(a) ${8{\times}6}$ cylinders at ${U/t=4.0}$
        and half-filling and  
        (b) ${16{\times}6}$ cylinders at ${U/t=8.0}$ and ${n=0.875}$.
		The MPS bond dimension is increased every other sweep 
		and is written alongside the corresponding data points.
		The solid lines are linear fits through the last five
        data points of each series 
		and indicate the zero-truncation-error extrapolations
        of the ground-state energies.
	}
	\label{fig:discarded_weight_extrapolation}
\end{figure}

After having compared the computational costs for fixed $m$,
we now investigate how the change of basis influences the block entropy
and the convergence of both methods.
Since the MPS bond dimension required to reach a fixed truncation error 
grows exponentially with the von Neumann entropy of the DMRG blocks, $S(i)$,
even small changes in the entropy can influence the convergence significantly. 
Figure~\ref{fig:blockentropy} shows 
comparisons of $S(i)$ in real and hybrid representations 
for the half-filled system at ${U/t=4.0}$, Fig.~\ref{fig:blockentropy}(a), 
and for the doped system, ${n=0.875}$, at ${U/t=8.0}$,
Fig.~\ref{fig:blockentropy}(b).
In Fig.~\ref{fig:blockentropy}(a), 
it can be seen that the block entropy in the
hybrid-space representation differs only slightly from that in the
real-space representation.
Note that if we cut the system between neighboring rings 
of the cylinder (gray lines),
the block entropy actually is the same in both cases, as expected.
This can also be seen in Fig.~\ref{fig:blockentropy}(b); however, here
the entropy in the hybrid representation is perceptibly higher between
these points.
This illustrates that the nonlocal nature of the interaction within
the rings does lead to a moderate increase of the entropy for cuts
within the rings in the hybrid representation, especially as ${U/t}$ is
increased and the system is doped.
Figure~\ref{fig:discarded_weight_extrapolation} illustrates
the good agreement of the ground-state energies for fixed $m$ as well
as after extrapolation to zero truncation error
for both parameter sets in Fig.~\ref{fig:blockentropy}. 
The slight divergence in the extrapolations for the
hybrid versus the real-space calculations, especially evident in
Fig.~\ref{fig:discarded_weight_extrapolation}(b), shows the limitations of the 
extrapolation, especially for the real-space results.
As can be clearly seen, 
the maximum MPS bond dimension available is significantly larger for
the hybrid-space algorithm,
resulting in a more well-controlled extrapolation and higher accuracy
in the extrapolated energy.
In Fig.~\ref{fig:discarded_weight_extrapolation}, we expect
that the real-space results would converge towards the hybrid-space
results if $m$ could be further increased in the real-space calculation;
this is not practically possible here.
For narrower systems, 
the energies converge more rapidly with $m$,
and both methods yield results that agree very well.
For example, for the $16{\times}4$ cylinder at $U/t=8.0$ and $n=0.875$,
we obtain the same extrapolated energy, $E_0/t=-0.75114(2)$,
for each method.

Even with the significant reductions in computational costs,
achieving good convergence for wider cylinders is still very expensive, 
making an efficient parallelization indispensable.
In this respect, the hybrid-space DMRG has an advantageous property:
the additional momentum quantum number leads to finer-grained 
quantum number sectors, which makes for better load balancing.
In particular, the momentum quantum numbers
yield equal-sized quantum number sectors, 
resulting in more equal-sized chunks of work.
In our implementation, we apply shared-memory parallelization  
to the combined loops over quantum number sectors and terms of the
Hamiltonian, as described for classical DMRG
in Ref.~\cite{hager2004parallelDMRG}.
In order to further extend the applicability of the hybrid algorithm,
additional steps such as real-space
parallelization~\cite{stoudenmire2013parallel} or efficient
distributed memory parallelization~\cite{chan2004parallel} could be
implemented.

\section{Results}
\label{sec:results}

We study the ground state of the doped Hubbard model 
on width-4 and width-6 cylinders for ${U/t=4.0}$ and ${U/t=8.0}$
at filling ${n=N^{-1}\sum_{{\bf r}\,\sigma} n_{{\bf r}\,\sigma}=0.875}$.
We choose these two values of $U/t$
for the following reasons:
(i) In general, we want to compare the physics and the performance of
the method at moderate coupling with that at strong coupling.
(ii) It is interesting to investigate the stability
of inhomogeneous phases as the interaction strength is changed, in
particular, whether a stripe phase remains stable for weaker interaction.
(iii) For real-space DMRG methods, the numerical convergence, i.e.,
the behavior of the block entropy, 
is generally more favorable at strong coupling.
(Real-space DMRG actually becomes exact in the atomic limit rather
than the strong-coupling limit, but a local interaction that is strong
relative to the hopping generally brings the system closer to that limit.)
As was discussed in Sec.~\ref{sec:measured_performance}, the
hybrid-space DMRG has the same block entropy as the real-space method
at cuts between cylinder rings and moderate excess entropy for cuts
within the rings (see Fig.~\ref{fig:blockentropy}), an excess which
becomes smaller for weaker interaction strength.
Therefore, the hybrid-space algorithm has slightly better relative
convergence at moderate interaction strength and can also handle a
larger bond dimension $m$ than the real-space method, so that the
moderate-interaction regime is more accessible than with the
real-space method.

In the following, we calculate the
ground-state energy, the site occupancy, 
and equal-time spin, charge, and pair-field correlation functions.
All results are extrapolated to zero truncation error.
The maximum MPS bond dimension was at least ${30\,000}$ for
cylinder lengths $16$ and $32$,
${27\,500}$ for length $48$, and ${25\,000}$ for length $64$.

\begin{figure}[htpb]
	\includegraphics[width=8.6cm]{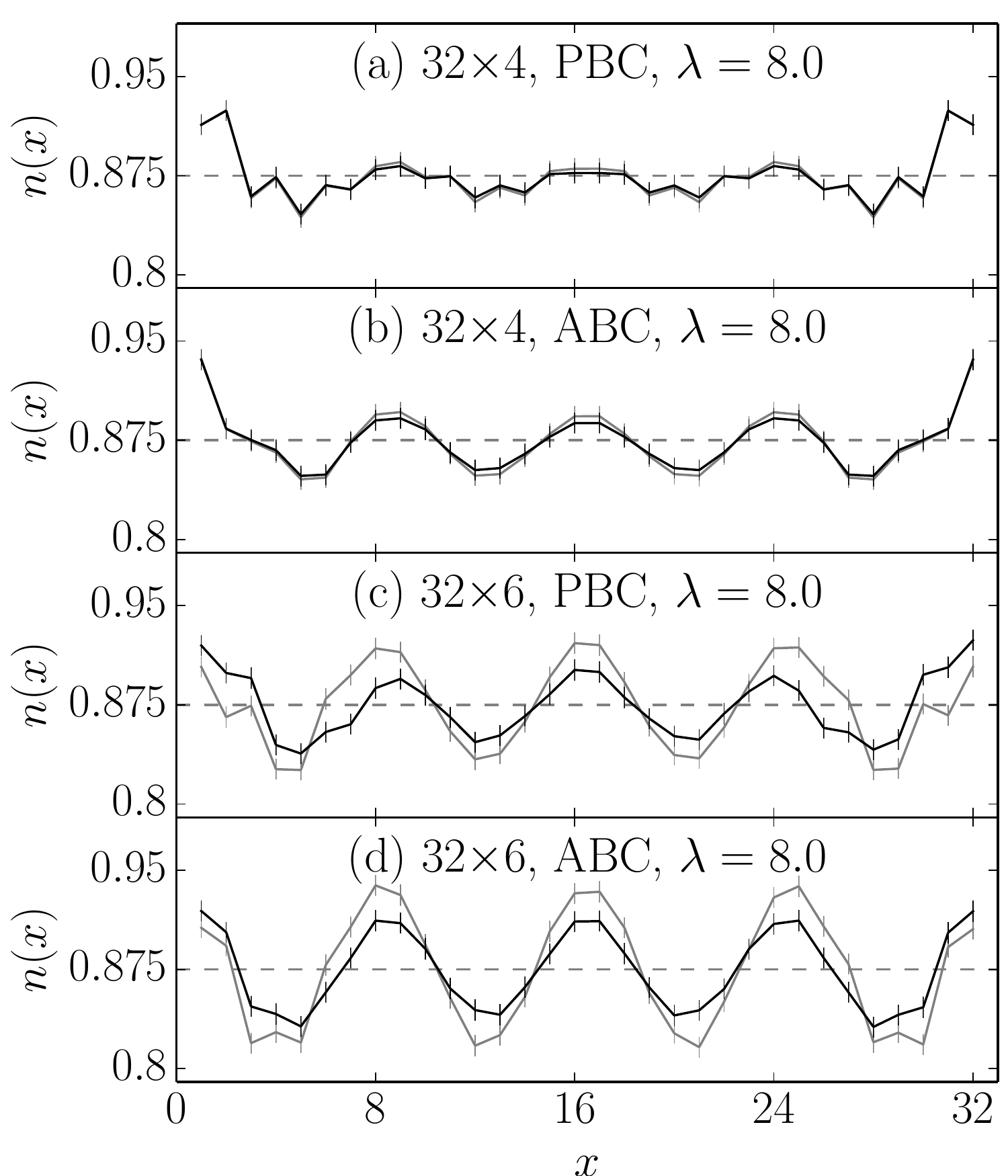}
	\caption{
		Charge-density distribution $n(x)$ for length-32 Hubbard cylinders
		at ${U/t=4.0}$ and ${n=0.875}$.
		Panels (a)--(d) show width-$4$ and width-$6$ cylinders with		
		periodic (PBC) and antiperiodic (ABC) transverse boundary conditions,
        as indicated.
		The black lines show the zero-truncation-error extrapolated densities,
		and the gray lines show the non-extrapolated values.
		The gray dashed lines indicate the average filling.
	}
	\label{fig:density_stripes}
\end{figure}
\begin{figure}[htpb]
	\includegraphics[width=8.6cm]{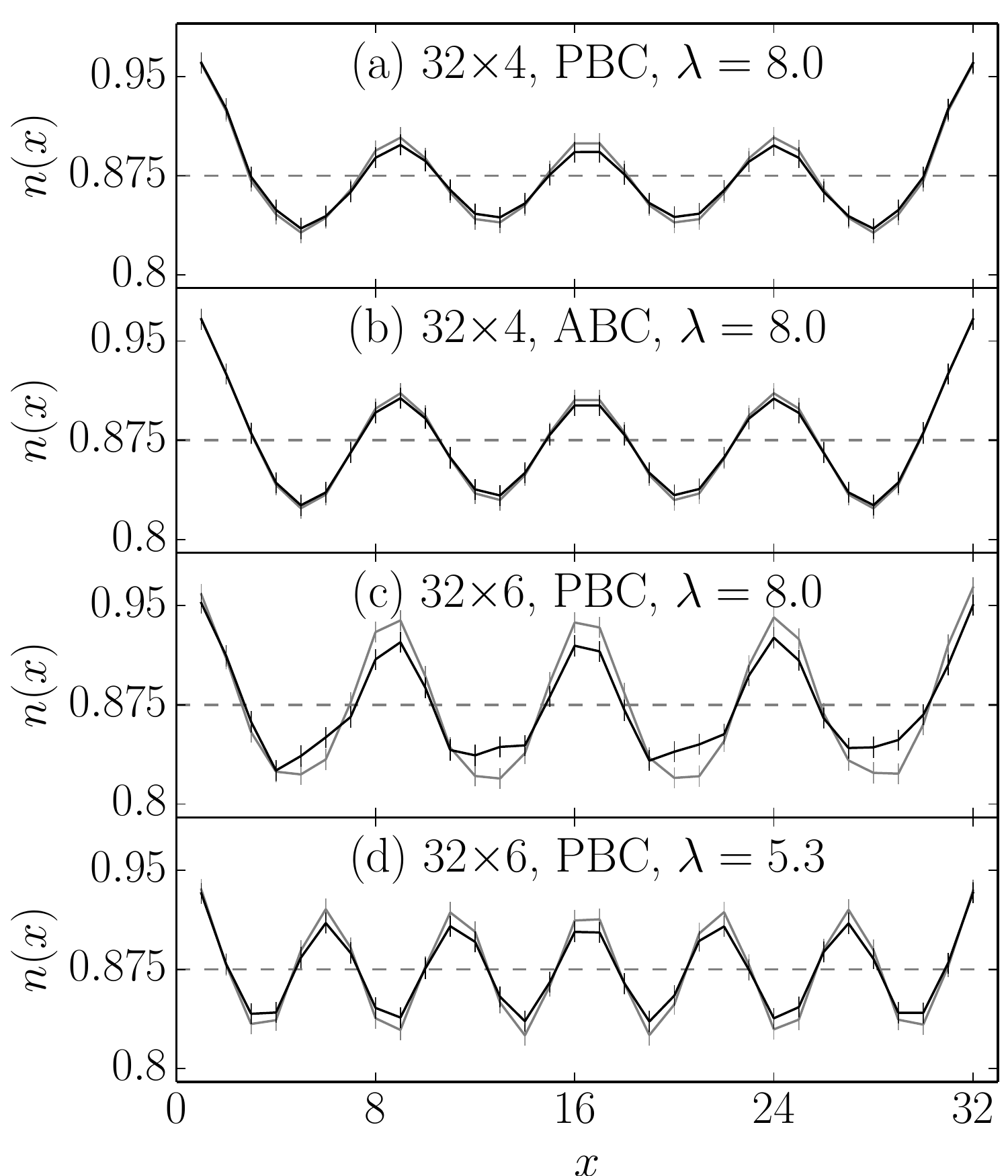}
	\caption{
		Charge-density distribution $n(x)$ for length-32 Hubbard cylinders
		at ${U/t=8.0}$ and ${n=0.875}$.
		Panels (a)--(d) 
		show cylinder with different width, boundary conditions,
		and wavelength $\lambda$ of the charge-density stripes,
		as indicated.
		The black lines show the zero-truncation-error extrapolated densities,
		and the gray lines show the non-extrapolated values.
		The gray dashed lines indicate the average filling.
	}
	\label{fig:density_stripes_u8}
\end{figure}
Figure~\ref{fig:density_stripes} shows the charge-density distribution
in the longitudinal direction, 
${n(x) = L_y^{-1} \sum_{y\,\sigma}n^{\vphantom{\dagger}}_{x\,y\,\sigma}}$,
for ${U/t=4.0}$ and different widths $L_y$ and boundary conditions.
In all cases, we find a stripe pattern with wavelength 8;
for $L_y=4$, each stripe contains $4$ holes
[Figs.~\ref{fig:density_stripes}(a)~and~\ref{fig:density_stripes}(b)],
and for $L_y=6$ each stripe contains $6$ holes
[Figs.~\ref{fig:density_stripes}(c)~and~\ref{fig:density_stripes}(d)].
For larger interaction strength, ${U/t=8.0}$,
Fig.~\ref{fig:density_stripes_u8}, we find two different
stripe configurations, wavelength ${\lambda=8.0}$
[Figs.~\ref{fig:density_stripes_u8}(a)--\ref{fig:density_stripes_u8}(c)] 
and wavelength ${\lambda=5.3}$ (more exactly, ${\lambda=16/3}$)
[Fig.~\ref{fig:density_stripes_u8}(d)].
For both configurations, we
add a pinning field, Eq.~(\ref{eqn:PinningField}), with the
appropriate wavelength to stabilize the state.
In most cases it is sufficient to add the pinning field during the
initial sweeps of the calculation
and ramp its amplitude $P$ down to zero in subsequent sweeps.
However, for width $6$ cylinders at $U/t=8.0$,
the amplitude of the field must be held finite (we take ${P=0.01}$)
during the entire calculation in order to stabilize the 
$\lambda=8.0$ stripe pattern.
After subtracting the field energy, we find that the ${\lambda=8.0}$ phase
is lower in energy and is thus globally stable; 
it should also be noted that 
the ${\lambda=5.3}$ phase only occurs for the ${L_y=6}$ lattices with periodic
boundary conditions.

\begin{figure}[htpb]
	\includegraphics[width=8.6cm]{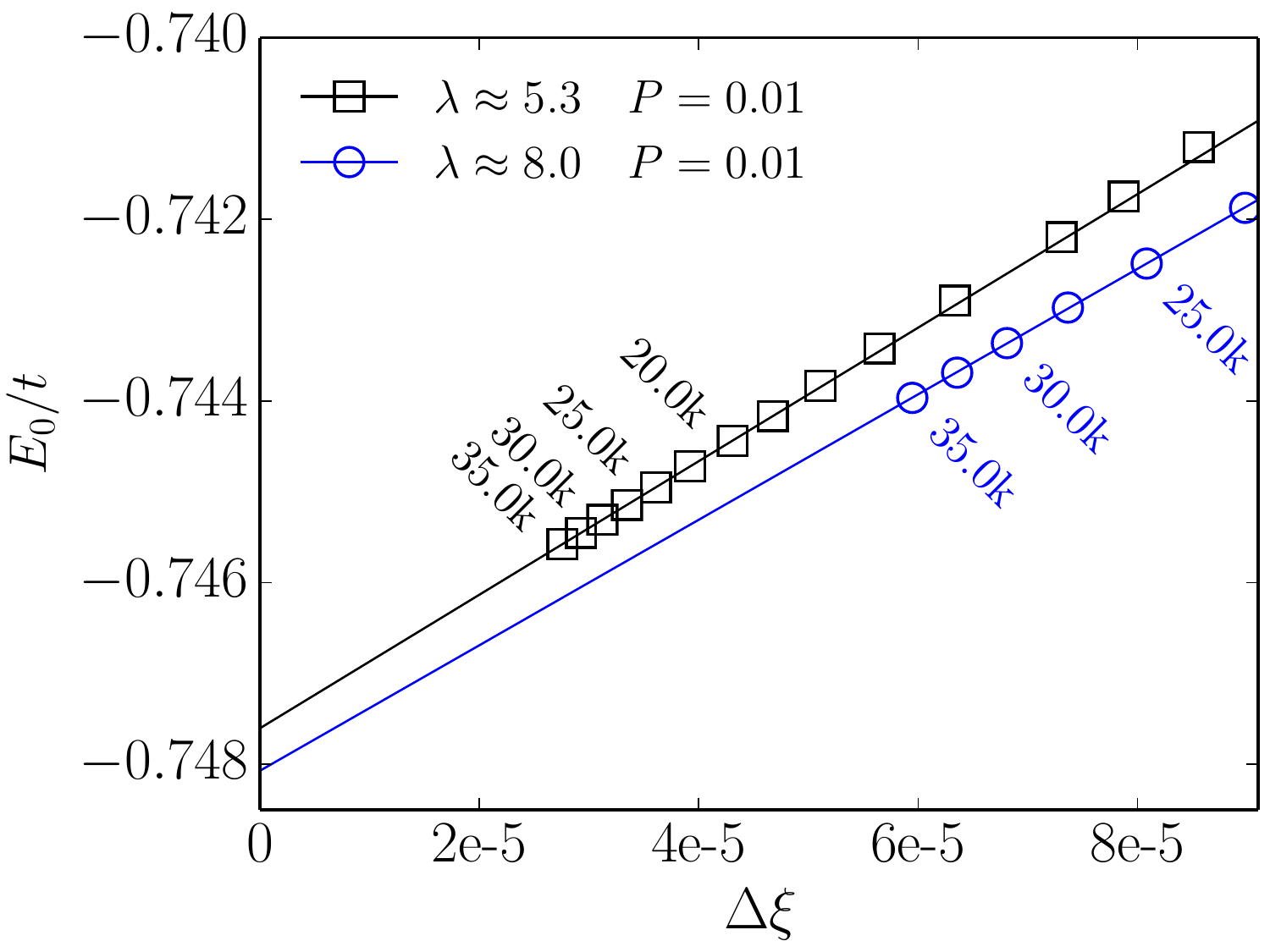}
	\caption{
		(Color online) Discarded weight extrapolation of the energy for
		$\lambda=5.3$ and $\lambda=8.0$ on ${16{\times}6}$ cylinders
		at $U/t=8.0$ and $n=0.875$.
		The energy contribution of the pinning field
		has been subtracted, and the MPS bond dimension $m$ 
		is indicated by the labels alongside selected data points.
		}
	\label{fig:lambda8vs5extrapolation}
\end{figure}

In order to elucidate the convergence of the $\lambda=5.3$  and $\lambda=8.0$
states for $L_y=6$ at $U/t=8.0$,
we examine the discarded-weight extrapolation of both
pinning-field-stabilized states in
Fig.~\ref{fig:lambda8vs5extrapolation}.
As can be seen, the $\lambda=8.0$ state
has lower energy for any fixed $\Delta\xi$ as well as for
$\Delta\xi\to 0.0$,
but the $\lambda=5.3$ state has lower energy for any accessible fixed $m$.
Because of this, the $\lambda=8.0$ state
becomes unstable if $P$ is set to zero at any point during the calculation
due to the fact that the DMRG algorithm is always driven towards the state 
that has the lowest energy within the finite reduced subspace of the Hilbert
space treated, i.e., the state space with fixed MPS bond dimension $m$.
The ground state within this subspace is not necessarily the ground state 
for fixed truncation error $\Delta\xi$ or, indeed, for $\Delta\xi\rightarrow0$.
In general, the DMRG should converge to the true ground state at some
level of accuracy (because it becomes exact for $m\rightarrow\infty$),
but the required $m$ may be inaccessible.
Thus, applying a pinning field is necessary here in order to 
enable the DMRG to track both states as the discarded weight is
varied; in practice, we use the fixed pinning field amplitude of $P=0.01$ for
the depicted range of $\Delta \xi$ for both states even though it would have been possible to set $P=0$ in
the latter part of the $\lambda=5.3$ calculation.

\begin{figure}
	\includegraphics[width=8.6cm]{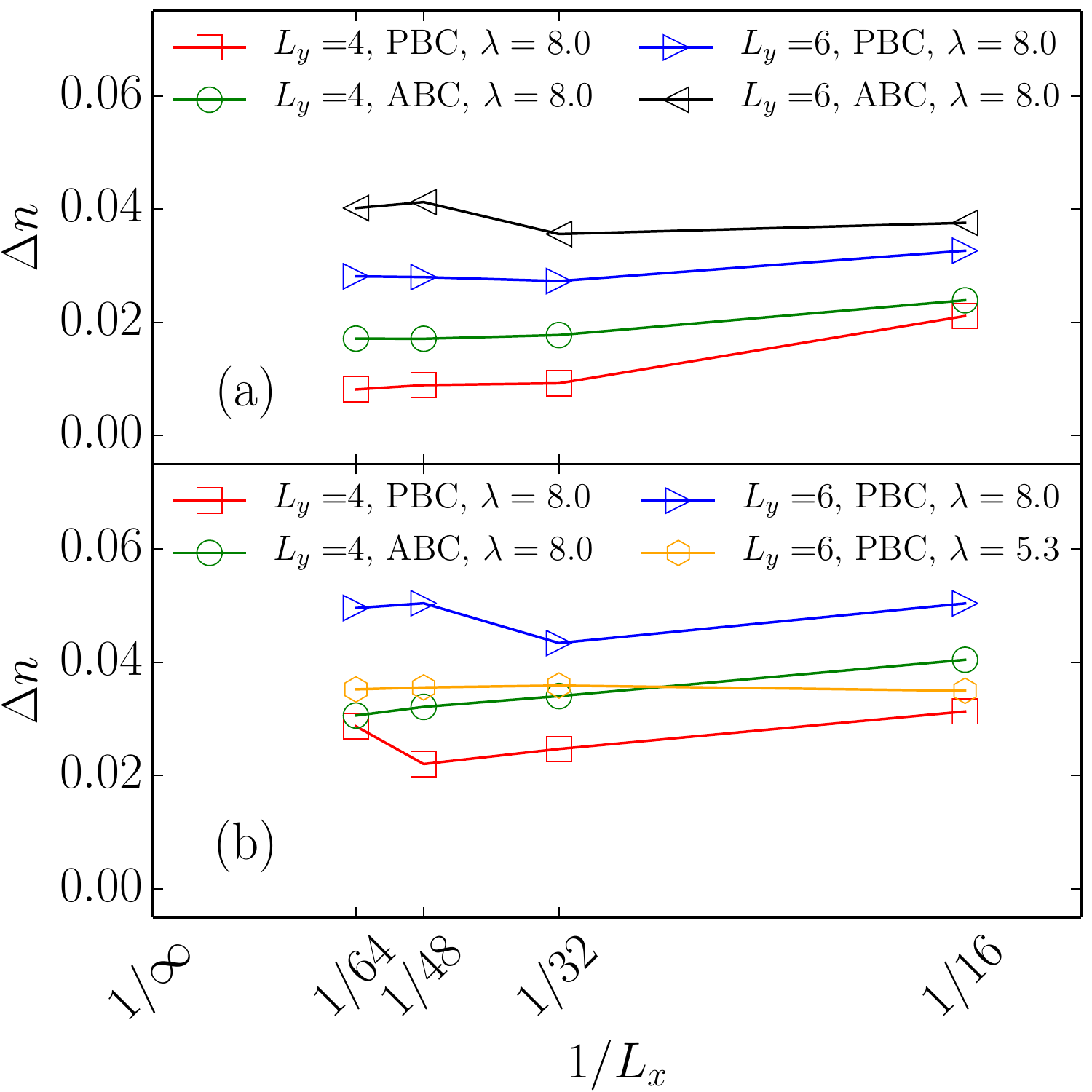}
	\caption{
		(Color online) Amplitude of the charge-density stripes,  ${\Delta n}$,
		for width-$4$ and width-$6$ cylinders
		with periodic (PBC) and antiperiodic (ABC) 
		transverse boundary conditions
		at ${n=0.875}$ with (a) ${U/t=4.0}$ and (b) ${U/t=8.0}$ 
        as a function of the inverse cylinder length, $1/L_x$.
	}
	\label{fig:density_stripes_amplitude}
\end{figure}
The amplitude of the charge-density modulations, 
${\Delta n = 1/2 \, \{\max_x[n(x)]-\min_x[n(x)]\} }$, 
is plotted as a function of the inverse cylinder length in
Fig.~\ref{fig:density_stripes_amplitude} for both values of $U/t$.
The trend shows that in all cases the stripes have finite amplitude in
the infinite-cylinder length limit.
The fact that the stripes are enhanced for ${L_y=6}$ 
indicates a striped ground state in the two-dimensional limit.
As the interaction is increased from ${U/t=4.0}$ to ${U/t=8.0}$, the
amplitude of charge density in the ${\lambda=8.0}$ stripes increases.

\begin{figure}
	\includegraphics[width=8.6cm]{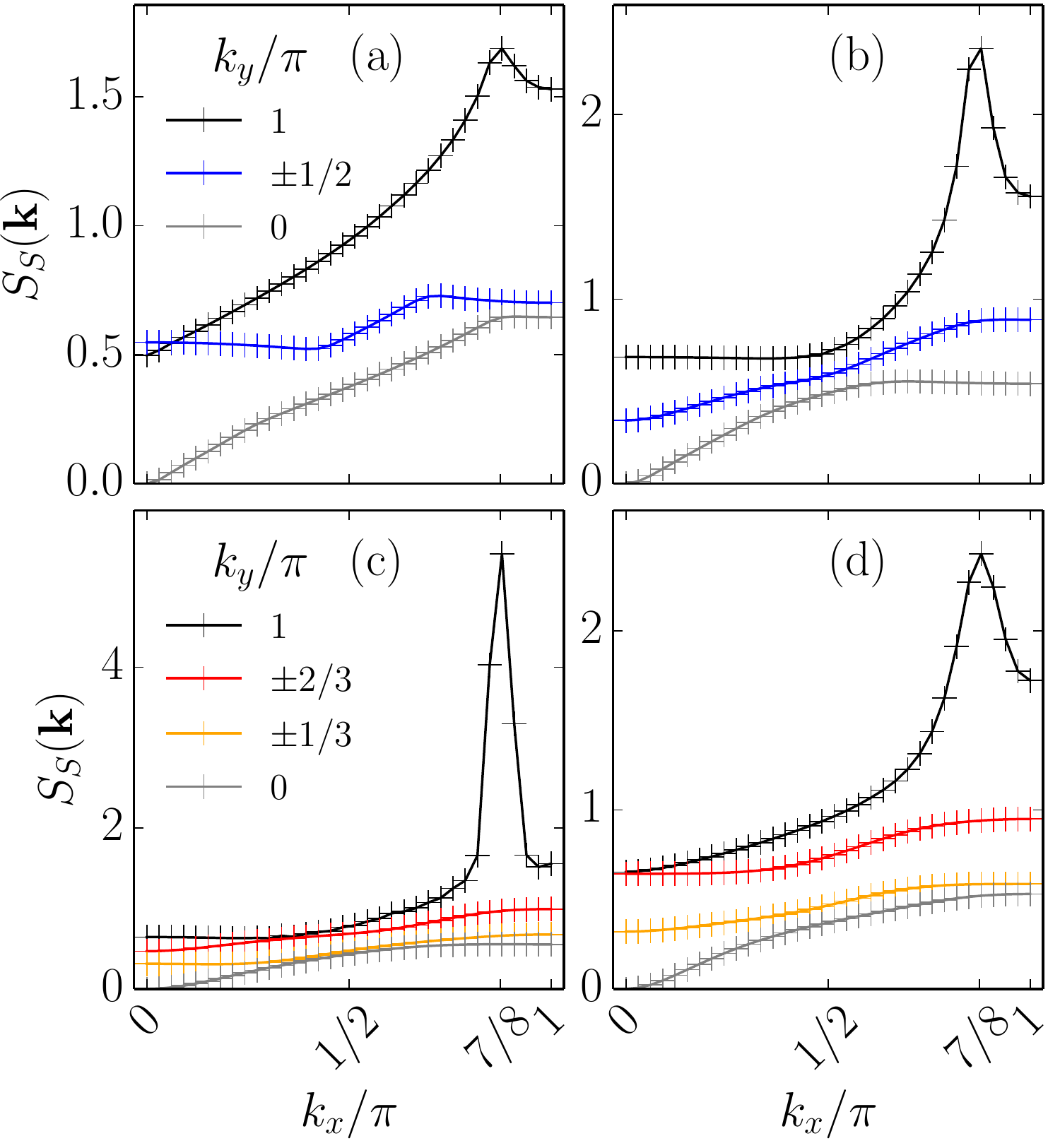}
	\caption{ 
		(Color online) 
		Spin structure factor ${S_S({\bf k})}$ for (a) and (b) ${32{\times}4}$ 
		and (c) and (d) ${32{\times}6}$  Hubbard cylinders 
		with periodic [(a) and (c)] 
		and antiperiodic [(b) and (d)] boundary conditions
		at ${n=0.875}$ and ${U/t=4.0}$.
	}	
	\label{fig:spin_structure}	
\end{figure}
\begin{figure}
	\includegraphics[width=8.6cm]{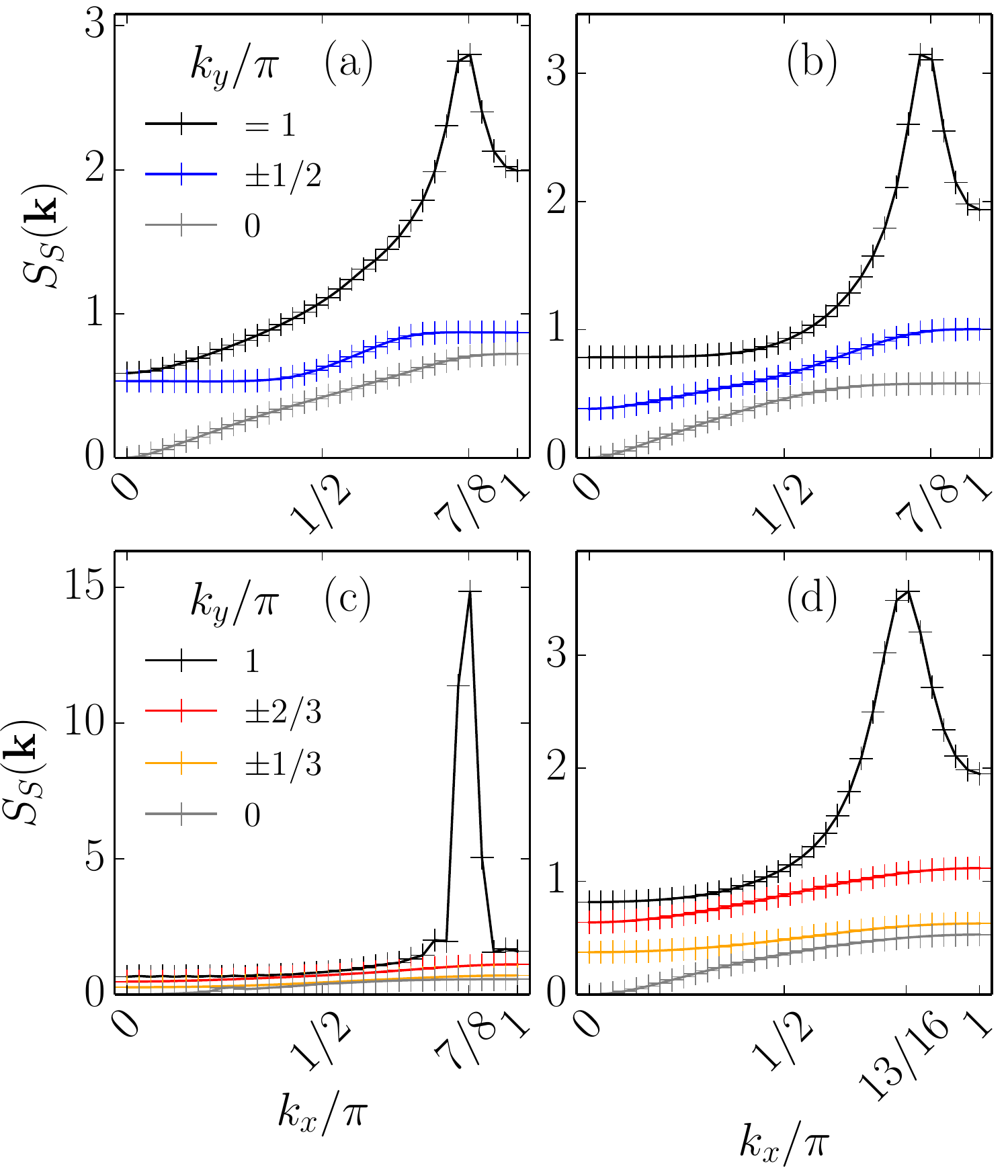}
	\caption{ 
		(Color online) Spin structure factor ${S_S({\bf k})}$
		at ${n=0.875}$ and ${U/t=8.0}$ for 
		(a) ${L_y=4}$, PBC, ${\lambda=8.0}$,
		(b) ${L_y=4}$, ABC, ${\lambda=8.0}$,
		(c) ${L_y=6}$, PBC, ${\lambda=8.0}$,
		and (d) ${L_y=6}$, PBC, ${\lambda=5.3}$.
	}	
	\label{fig:spin_structure_U8}	
\end{figure}
Typically, a striped charge-density distribution in the doped Hubbard
model is accompanied by a striped staggered
spin-density distribution with a wavelength that is double 
that of the charge-density distribution~\cite{white2003StripesHubbard}.
Measuring these stripes in hybrid space would require breaking the
translation invariance in the transverse direction,
which could only be done at the cost of slowing down the algorithm
significantly.
Instead, we calculate the structure factor of the equal-time spin correlations,
\begin{equation}
	S_{S}({\bf k}) = 
	\frac{1}{N} \sum_{\bf{r} \, {\bf r}'} 
	e^{\rm{i} {\bf k} \, ({\bf r} - {\bf r}')} \;
	S({\bf r},{\bf r}') \, ;
\end{equation}
for finite cylinder length we expand in the particle-in-a-box eigenmodes,
as in Ref.~\cite{hager2005stripe}.
Figures~\ref{fig:spin_structure} and~\ref{fig:spin_structure_U8}
show ${S_{S}({\bf k})}$ for ${U/t=4.0}$ and ${U/t=8.0}$, respectively, at
${n=0.875}$
(the same parameter sets as in 
Figs.~\ref{fig:density_stripes} and~\ref{fig:density_stripes_u8}).  
As can be seen in 
Figs.~\ref{fig:spin_structure}(a)--\ref{fig:spin_structure}(d) for
${U/t=4.0}$ and in 
Figs.~\ref{fig:spin_structure_U8}(a)--\ref{fig:spin_structure_U8}(c) for
${U/t=8.0}$, 
${S_{S}({\bf k})}$ is peaked at 
${{\bf k}\approx(\pm7/8\,\pi,\pi)}$ in all cases
in which period ${\lambda=8.0}$ stripes are present.
Thus, the spin correlations are antiferromagnetic with a $\pi$-phase
shift every eighth site, as expected.
For the ${\lambda = 5.3}$ stripes, Fig.~\ref{fig:spin_structure_U8}(d),
the peak shifts to ${{\bf k}\approx(\pm13/16\,\pi,\pi)}$
which corresponds to antiferromagnetic spin correlations with a $\pi$-phase
shift every $5.3$ sites, also compatible with the stripe structure.
As we have seen for the amplitude of the charge-density stripes,
the amplitudes of the peaks of the spin structure factor
also increase with increasing interaction strength.

\begin{figure}[htpb]
	\includegraphics[width=8.6cm]{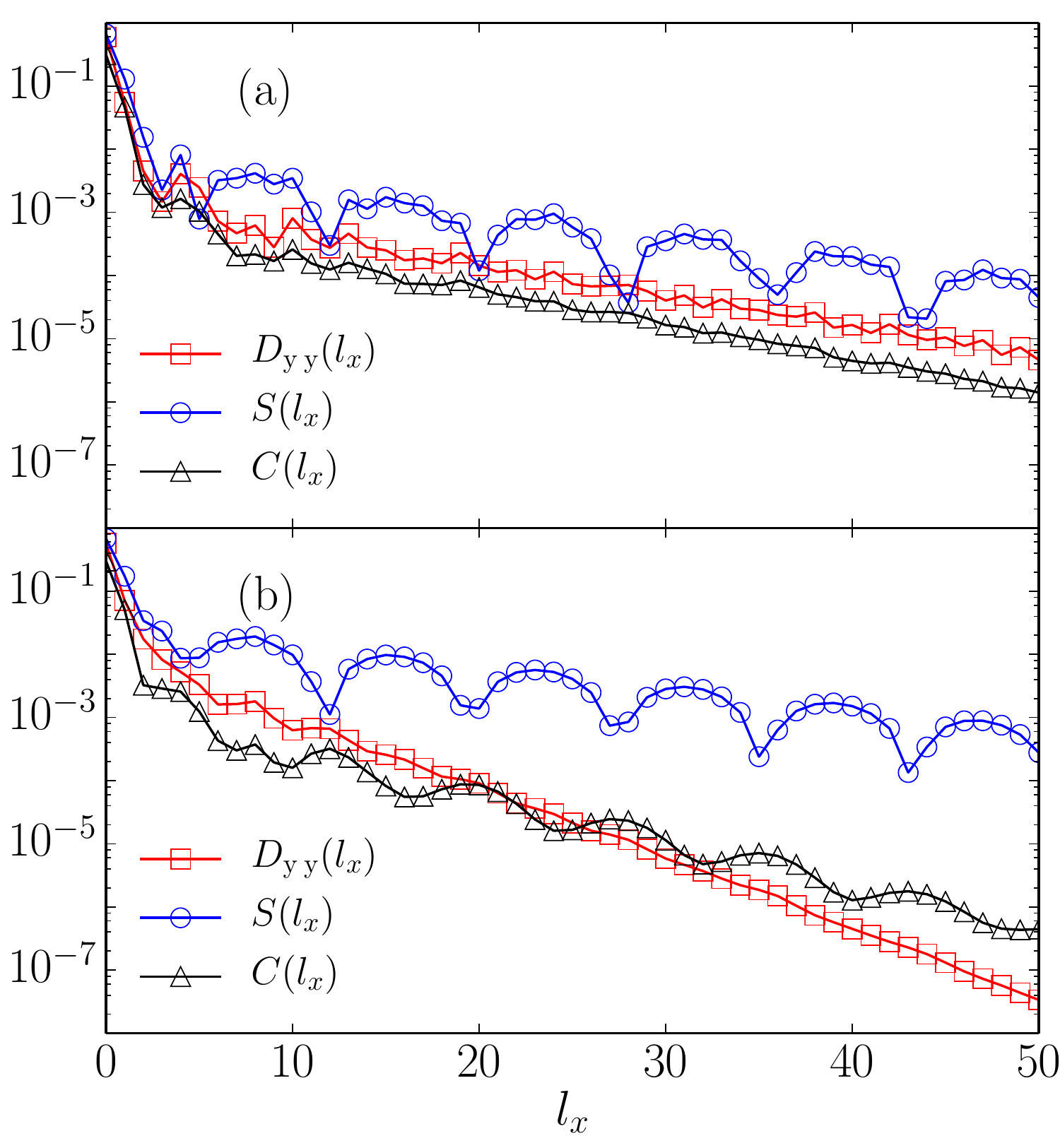}
	\caption{
		(Color online) 
		Equal-time pair-field ${D_{\rm y\,y}(l_x)}$, spin ${S(l_x)}$, 
		and charge ${C(l_x)}$ correlations
		as a function of the longitudinal distance $l_x$ 
		for ${64{\times}4}$ Hubbard cylinders
		at ${n=0.875}$ and ${U/t=4.0}$ for (a) periodic 
		and (b) antiperiodic boundary conditions.
	}
	\label{fig:correlations}	
\end{figure}

In order to test for ${d_{x^2-y^2}}$-pairing-induced superconductivity, 
we have calculated the equal-time pair-field correlation functions
and have compared their decay to that of the spin and charge correlations.
To compensate for the stripe structure of the ground state,
we take averaged absolute values; e.g., for the spin correlation
functions we define
\begin{equation}
	S(l_x) = 
	\frac{1}{8} \sum_{x=(L_x-l_x)/2-4}^{(L_x-l_x)/2+3} |S(x,x+l_x)| \, .
\end{equation}
The decay of the correlations with distance along the cylinder 
is shown in Fig.~\ref{fig:correlations} for width-4 cylinders. 
As can be seen by the approximately linear behavior of the
envelopes of the correlation functions on the semilogarithmic scale,
all three correlation functions decay exponentially at moderate to
long distances for both periodic and antiperiodic boundary conditions.
For periodic boundary conditions, all three correlation functions decay
approximately at the same rate,
with the spin correlations having a larger absolute value.
For antiperiodic boundary conditions, the spin correlations are
dominant for larger distance.
Thus, we do not find that pairing correlations are 
long-range or, indeed, even dominant.
\begin{figure}[htpb]
	\includegraphics[width=8.6cm]{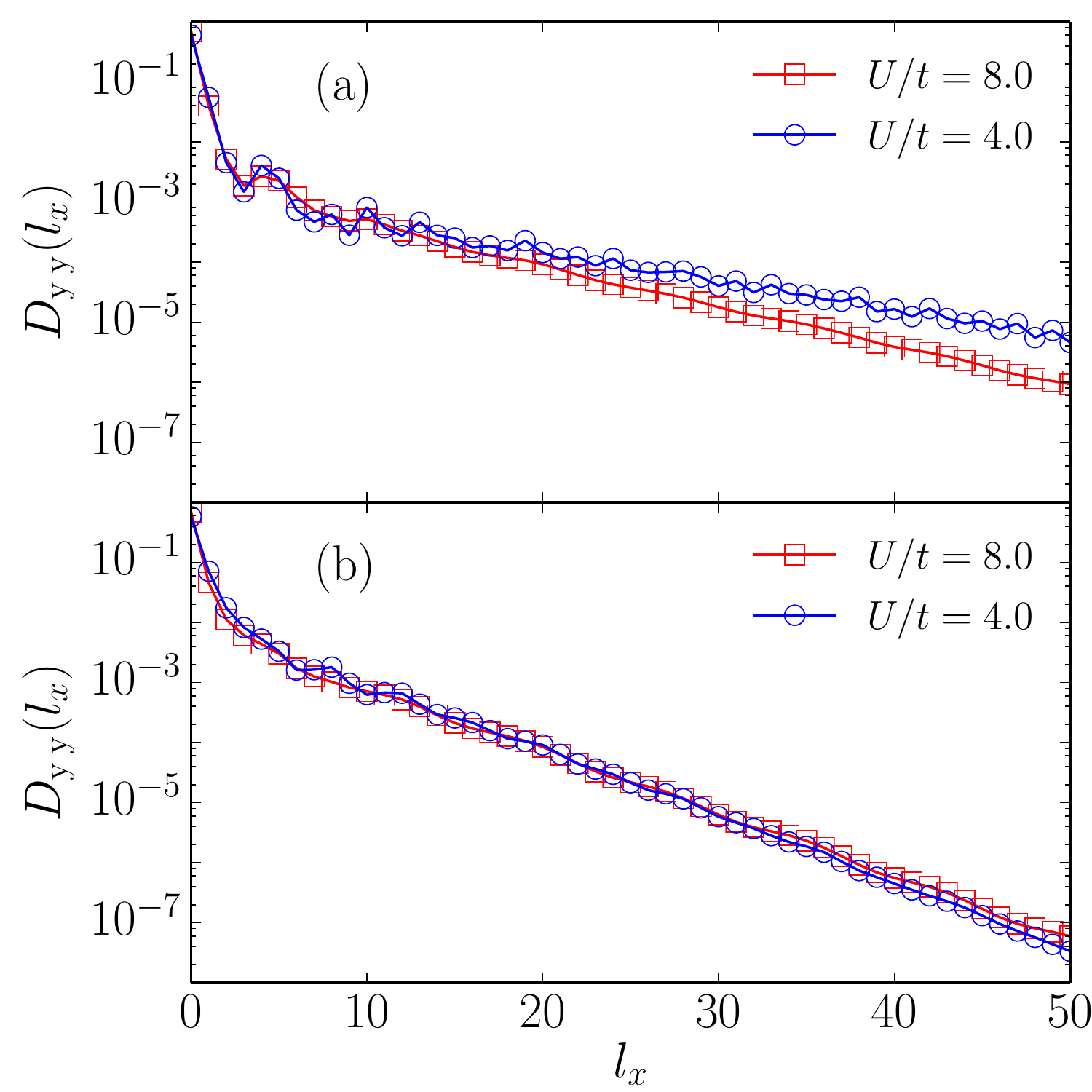}
	\caption{
		(Color online) 
		Equal-time pair-field correlation functions
		for ${64{\times}4}$ Hubbard cylinders
		at ${n=0.875}$ for different $U/t$ for (a) periodic 
		and (b) antiperiodic boundary conditions.
	}
	\label{fig:pair_field_correlations}	
\end{figure}

We investigate the effect of interaction strength on the
strength of pairing correlations by comparing the pair correlation
functions for the ${\lambda=8.0}$ stripes for ${U/t=4.0}$ and ${U/t=8.0}$ in
Fig.~\ref{fig:pair_field_correlations}.
For periodic boundary conditions,
Fig.~\ref{fig:pair_field_correlations}(a), there is a slight
suppression of the exponentially decaying correlation function as $U/t$
is increased, whereas for antiperiodic boundary conditions, 
Fig.~\ref{fig:pair_field_correlations}(b),
changing $U/t$ has virtually no effect.

The obvious question to be addressed is why the pair correlations,
and, indeed, also the charge and spin correlations, are strong at
short length scales, but have no long-range or quasi-long-range (i.e.,
critical power-law) order.
Here we point out that the stripe state breaks the translational
symmetry and is locked into the finite lattice, so that the charge
order manifests itself in a variation of the local charge density.
One then expects the charge correlation 
{\em with the local order subtracted out} 
to be short-range, i.e., exponentially decaying.
A similar argument holds for the spin correlations, which are locked
to the charge correlations (but have twice the wavelength and a 
$\pi$-phase shift between stripes).
The $d_{x^2-y^2}$ pair correlations do show strong short-range
correlations, but are also exponentially decaying, meaning that
pairing is not even present in the one-dimensional sense, i.e., that
the correlations decay as a power law.
To explain this behavior, we make three points:
(i) The locked-in charge order could preclude any other than
short-range order in all
correlation channels.
(ii) The pairing correlations are measured perpendicularly to the
stripes, i.e., in a direction in which charge transport for static
stripes would not be expected.
(Note that the cylindrical geometry locks in transverse stripes, and that
measurement of pair correlations in the transverse direction over any
significant length is not possible due to limitations in the treatable
system width.)
(iii) For this doping of the system and wavelength of the stripes, the
stripes are completely filled with holes and thus insulating, so that
quasi-long-range pair correlations would not be expected.
In addition, it is fair to point out that, as earlier works on two-leg
ladders have shown~\cite{noack1997HubbardLadder,dolfi2015HubbardLadder},
convergence of the long-range part of the pairing correlations occurs
very slowly as $m$ is increased in the DMRG; 
an MPS bond dimension of ${30\,000}$ states might still be
insufficient to restore algebraic decay over long length scales for
width-4 cylinders.

Finally, we obtain accurate estimates of the ground-state energies 
in the infinite-length-cylinder-limit by carrying out consecutive
extrapolations in the discarded weight, followed
by the inverse cylinder length.
\begin{figure}[htpb]
	\includegraphics[width=8.6cm]{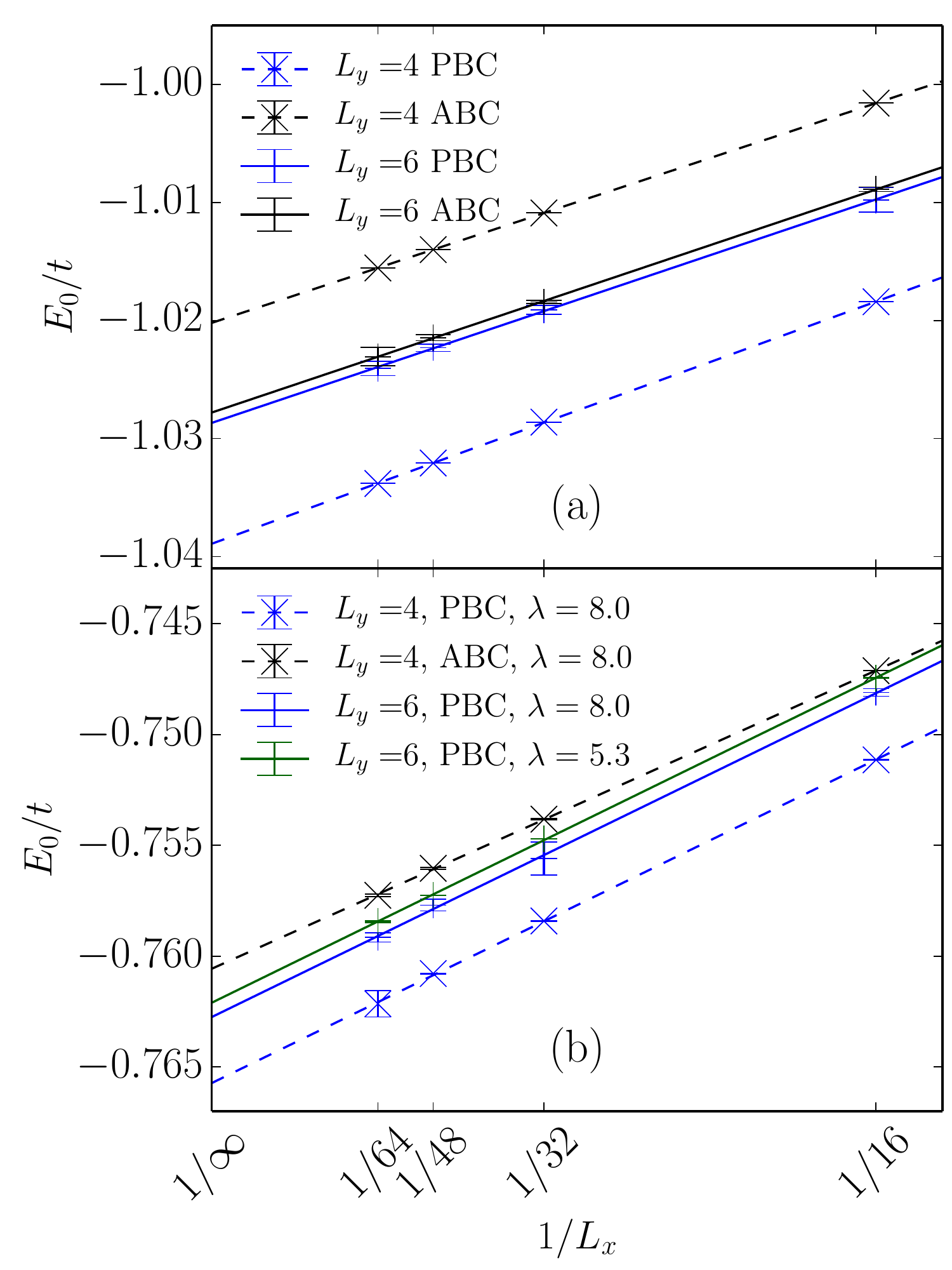}
	\caption{
	(Color online)
		Ground-state energies 
		for doped width-$4$ and width-$6$ Hubbard cylinders 
		with periodic (PBC) and antiperiodic (ABC) 
		transverse boundary conditions 
		at ${n=0.875}$ and (a) ${U/t=4.0}$ and (b) ${U/t=8.0}$
		as a function of the inverse cylinder length.
        In (b), the wavelength $\lambda$ of the charge-density stripe 
        is indicated.
		The energies are extrapolated to zero discarded weight, 
		with the estimated error from the extrapolation 
		indicated by the error bars.  
		The dashed and the solid lines 
		show the extrapolation to infinite cylinder length
		for width-$4$ and width-$6$ cylinders, respectively.
	}
	\label{fig:cylinder_length_extrapolation}
\end{figure}
The extrapolation to infinite cylinder length 
for ${U/t=4.0}$ and ${U/t=8.0}$ is shown in
Fig.~\ref*{fig:cylinder_length_extrapolation}.
Energies for cylinder lengths $16$, $32$, $48$, and $64$ fall almost
perfectly onto the linear fits
for all curves in both 
Figs.~\ref*{fig:cylinder_length_extrapolation}(a) 
and~\ref*{fig:cylinder_length_extrapolation}(b).
As can be seen, the error bars, which result from the 
discarded-weight extrapolations, get larger for wider cylinders,
but the data points still allow for a well-controlled and accurate
extrapolation.
Note that in Fig.~\ref*{fig:cylinder_length_extrapolation}(b) 
the meta-stable state with a stripe wavelength ${\lambda=5.3}$ is
very close in energy to the ${\lambda=8.0}$ state.
Even though the effects of the discarded-weight extrapolation,
Fig.~\ref{fig:lambda8vs5extrapolation}, and the system-length
extrapolation can be larger than the energy difference, the energy
of the ${\lambda=8.0}$ state is consistently lower than that of the
${\lambda=5.3}$ state, both at finite
system length after the discarded-weight extrapolation and after
subsequently extrapolating to the
infinite-system-length limit.
The energies obtained for all parameters are given 
in Appendix~\ref{app:energies} 
in Tables~\ref{tab:energiesU4} and \ref{tab:energiesU8}.

\section{Summary and discussion}
\label{sec:conclusion}

We have investigated the applicability and usefulness of the DMRG in a
hybrid--real-momentum-space formulation for the two-dimensional Hubbard
model on cylindrical lattices.
We have first compared the computational costs of the
real-space and hybrid-space DMRG as a function of the system width in
theory and practice, and have shown the hybrid-space variant to
be significantly faster.
In particular, we have shown that, due; to conservation of the
additional momentum quantum number, 
the computational and memory costs of the hybrid-space DMRG are
essentially independent of the width of the cylinder 
for fixed dimension of the MPS bonds, $m$.
In practice, we have found that the computational cost is almost
width-independent and have obtained speedup factors of roughly $12$, $20$,
and $26$ for cylinder width $4$, $6$, and $8$ relative to the
real-space DMRG for fixed  $m$.
Subsequently, we have demonstrated,
that the entropy in the system is not increased 
significantly in the hybrid representation,
giving both formulations comparable convergence with  $m$.
Thus, we have shown that the hybrid-space ansatz widens
the applicability of the DMRG to Hubbard-like models,
especially for larger system widths.

We have then used the hybrid-space DMRG to investigate the
static properties of the ground state of the doped two-dimensional
Hubbard model at filling ${n=0.875}$ and interaction strengths 
${U/t=4.0}$ and ${U/t=8.0}$;
examining width-$4$ and width-$6$ cylinders, we have found that the
system forms a striped charge-density distribution of wavelength
${\lambda=8.0}$ for both values of $U/t$.
The magnetic ordering of the ground state is antiferromagnetic 
with a modulation of wavelength $16$.
For width~6 cylinders with periodic boundary conditions
and interaction strengths ${U/t=8.0}$,
we have also found a metastable ${\lambda=5.3}$ state,
which we have shown to be slightly higher in energy
for both finite cylinder length and in the infinite-length limit.
Furthermore, we have calculated pairing, spin, and charge correlation functions
and have found that the spin correlations have the slowest decay with
distance, while the pair-field and charge correlations have a
sub-dominant decay of comparable strength.
The behavior of the correlations is nearly identical 
for ${U/t=4.0}$ and ${U/t=8.0}$.

We now compare our results to those
of other methods,
first for ${U/t=4.0}$:
unsurprisingly, 
the ground-state energy we have obtained
is in excellent agreement 
with recent real-space DMRG calculations;
Leblanc \textit{et al}.\ reported a value of
${E_0/t=-1.028}$~\cite{leblanc2015HubbardBenchmark},
for infinite width-6 cylinders averaged over periodic and
antiperiodic boundary conditions, 
while we have obtained ${E_0/t=-1.0282(4)}$.
Recent calculations using the density matrix embedding
theory (DMET)~\cite{chan2012DMET} find a phase diagram
in which the system lies just within a superconducting phase for
${n=0.875}$ and ${U/t=4.0}$.
This phase is in very close proximity to both an antiferromagnetic and
to a metallic phase;
see Fig.~2 in Ref.~\cite{chan2016HubbardPhaseDiagram}.
In particular, the authors find
two states with somewhat different magnetic orderings and strengths of
$d$-wave pairing correlations that are energetically
very close to one another:
an incommensurate antiferromagnet 
for ${8{\times}2}$ clusters with ${E_0/t=-1.0288}$ 
and a homogeneous antiferromagnetic state 
for ${4{\times}4}$ clusters with ${E_0/t=-1.033(2)}$.
A recent DMET calculation on a ${16{\times}2}$
cluster~\cite{chanPrivateCommunication},
which is large enough to encompass the $\pi$-phase shift in spin
correlations between stripes, yields a lower energy, ${E_0/t=-1.0327(1)}$,
for the incommensurate antiferromagnet.
This improved energy for the incommensurate state is within error
bounds of that of the homogeneous antiferromagnetic state, leaving the
nature of the ground state within the DMET undetermined.
The improved energy for the incommensurate antiferromagnetic state and
for the homogeneous state agree well
with our results; however, we find only a incommensurate striped
state, no homogeneous state, in our calculations.
We note that we have been able to take larger cluster sizes into
account and that an incommensurate striped state is excluded for the
${4{\times}4}$ cluster in the DMET calculations.
Therefore, we find it likely that the ground state for this parameter
set indeed has an inhomogeneous, i.e., striped, magnetic order in
the thermodynamic limit.
Earlier DMRG studies of doped width-6 Hubbard cylinders found that
the charge-density stripes disappear in the ${L_x\rightarrow\infty}$
limit for ${U/t=3.0}$ and ${n\approx0.905}$~\cite{hager2005stripe}.
According to the phase diagram of the DMET calculations, this
parameter set falls just on the other side of the phase transition line
to a homogeneous metallic state.
Thus, these findings are not inconsistent with our results
or the DMET results.

The main additional feature of the results for ${U/t=8.0}$ is the presence of a
metastable higher energy state with stripe order of wavelength
${\lambda=5.3}$ in addition to the ${\lambda=8.0}$ ground state, found only
in the width-6 lattices with periodic boundary conditions.
The energy difference between these two states can only be resolved by
careful extrapolation in truncation error and system length.
We note, however, that these results are consistent with those of
other methods, including real-space 
DMRG, DMET, iPEPS~\cite{jordan2008iPEPS,corboz2014iPEPSstripes}, and
constrained-path auxiliary-field quantum Monte Carlo, as
discussed extensively in Ref.~\cite{zheng2017stripes}.
Thus, there is strong evidence 
that the ${\lambda=8.0}$ stripes are a robust feature of the
ground state of the two-dimensional Hubbard model at ${n=0.875}$ for
interaction strengths from ${U/t=4.0}$ to ${U/t=8.0}$.
 
\FloatBarrier
\begin{acknowledgements}
	We thank C.-M.\ Chung for useful discussions.
	G.E.\ and R.M.N.\ acknowledge support from 
	the Deutsche
	Forschungsgemeinschaft (DFG) through grant
	no.\  NO 314/5-2 in Research Unit FOR 1807.
    S.R.W.\ was supported by the U.S.\ National Science Foundation
    through grant no.\ DMR-1505406.
\end{acknowledgements}

\appendix

\section{MPO matrices}
\FloatBarrier
\label{app:mpo}

\newcommand\scalemath[2]{\scalebox{#1}{\mbox{\ensuremath{\displaystyle #2}}}}

The individual matrices of the MPO for the 
hybrid-space Hamiltonian~(\ref{eqn:HybbardTX})--(\ref{eqn:HybbardU})
are relatively large,  and their size and structure depend on the 
width of the cylinder, the mapping between the lattice and the MPO, 
and the position within the MPO in a nontrivial way.
Therefore, it is not possible to give the explicit form of the matrices
in a compact way.
Instead, we give stepwise instructions on how to construct the
explicit form of the MPO. 

MPO matrices are best written down as matrices of local operators.
For the case of spinful fermions 
with local basis ${(0,\uparrow,\downarrow,\uparrow\downarrow)}$, we use
\begin{flalign}
	\qquad {\bf c}^{\dagger}_{\uparrow}  = & 
	\left[
	\begin{matrix}
	0 & 0 & 0 & 0 \\
	1 & 0 & 0 & 0 \\
	0 & 0 & 0 & 0 \\
	0 & 0 & 1 & 0 
	\end{matrix}
	\right] \, ,
	\quad &
	{\bf c}^{\phantom{\dagger}}_{\uparrow}  = & 
	\left[
	\begin{matrix}
	0 & 1 & 0 & 0 \\
	0 & 0 & 0 & 0 \\
	0 & 0 & 0 & 1 \\
	0 & 0 & 0 & 0 
	\end{matrix}
	\right] \, ,
	\nonumber \\
	{\bf c}^{\dagger}_{\downarrow}  = & 
	\left[
	\begin{matrix}
	0 & 0 & 0 & 0 \\
	0 & 0 & 0 & 0 \\
	1 & 0 & 0 & 0 \\
	0 &-1 & 0 & 0 
	\end{matrix}
	\right] \, ,
	\quad &
	{\bf c}^{\phantom{\dagger}}_{\downarrow}  = & 
	\left[
	\begin{matrix}
	0 & 0 & 1 & 0 \\
	0 & 0 & 0 &-1 \\
	0 & 0 & 0 & 0 \\
	0 & 0 & 0 & 0 
	\end{matrix}
	\right] \, ,
	 \\
	{\bf 1}  = &  
	\left[
	\begin{matrix}
	1 & 0 & 0 & 0 \\
	0 & 1 & 0 & 0 \\
	0 & 0 & 1 & 0 \\
	0 & 0 & 0 & 1 
	\end{matrix}
	\right] \, ,
	\quad &
	{\bf F}  = &  
	\left[
	\begin{matrix}
	1 & 0 & 0 & 0 \\
	0 &-1 & 0 & 0 \\
	0 & 0 &-1 & 0 \\
	0 & 0 & 0 & 1 
	\end{matrix}
	\right] \,  \qquad
	\nonumber 
\end{flalign}
for the creation, annihilation, identity, and fermionic-sign operators.
Here we have incorporated the fermionic sign that occurs 
when the MPO is applied to an MPS through 
the ${\bf F}$ operator and the minus signs in 
${{\bf c}^{\dagger}_{\downarrow}}$ 
and ${{\bf c}^{\phantom{\dagger}}_{\downarrow}}$,
which are chosen according to the normal ordering   
${(
  {c^\dagger_\uparrow},
  {c^{\vphantom{\dagger}}_\uparrow},
  {c^\dagger_\downarrow},
  {c^{\vphantom{\dagger}}_\downarrow}
  )}$.

To familiarize the reader with the construction of the fermionic MPO,
we start by giving the form of the MPO for the one-dimensional Hubbard
model in real space:
\begin{equation}
	\def\arraystretch{1.3}
	W^{[i]}=
	\left[
	\begin{matrix}
		{\bf 1} & 0 & 0 & 0 & 0 & 0 \\ 
		t \, {\bf c}^{\vphantom{\dag}}_{\uparrow} & 0 & 0 &  0 & 0 & 0 \\ 
		-t \, {\bf c}^{{\dag}}_{\uparrow} & 0 & 0 &  0 & 0 & 0 \\ 
		t \, {\bf c}^{\vphantom{\dag}}_{\downarrow} & 0 & 0 & 0 & 0 & 0 \\ 
		-t \, {\bf c}^{{\dag}}_{\downarrow} & 0 & 0 & 0 & 0 & 0 \\ 
		U \, {\bf n}_{\uparrow} \, {\bf n}_{\downarrow} &
		{\bf c}^{{\dag}}_{\uparrow} {\bf F}&
		{\bf c}^{\vphantom{\dag}}_{\uparrow} {\bf F}&		
		{\bf c}^{{\dag}}_{\downarrow} {\bf F}& 
		{\bf c}^{\vphantom{\dag}}_{\downarrow} {\bf F}& 
		{\bf 1}
	\end{matrix}
	\label{eqn:realMPO}
	\right] \, ,
\end{equation} 
with 
${{\bf n}^{\vphantom{\dag}}_{\sigma}=
{\bf c}^{\vphantom{\dag}}_{\sigma} \,
{\bf c}^{{\dag}}_{\sigma}}$.
Here, $W^{[i]}$ is the matrix for the $i$-th site,
and the row and column indices correspond to the virtual MPO bonds.
Within each virtual bond, 
the first and the last channels
are used as ``initial'' and ``target'' channels of the MPO, 
and within these channels, identity operators
establish connections through the entire MPO from left to right.
The on-site repulsion is local in real space 
and thus its operator directly connects the initial and target
channels on each site.
For the hopping term, four additional channels are
needed to connect the appropriate combinations of 
creation and annihilation operators on neighboring sites.
To complete the MPO, the matrix for the first (last)
site must be multiplied with the column (row) 
unit vector ${\bf e}_6^T$ (${\bf e}_1^{\vphantom{T}}$).

Using the same basic concepts as in Eq.~(\ref{eqn:realMPO})
we can now construct the MPO for the first two parts~(\ref{eqn:HybbardTX})
and~(\ref{eqn:HybbardTY}) of the hybrid-space Hamiltonian,
i.e., the MPO for the hopping terms.
For a width-2 cylinder, 
the MPO matrices can be written as
\begin{align}
	\def\arraystretch{1.3}
	W^{[i]}= 
	\left[{
	\begin{array}{c!{\color{gray}\vrule}cccc!{\color{gray}\vrule}cccc!{\color{gray}\vrule}c} \arrayrulecolor{gray}
		{\bf 1} & 
		0 & 0 & 0 & 0 & 0 & 0 & 0 & 0 & 0 \\ 
		\arrayrulecolor{gray} \hline
		t \, {\bf c}^{\vphantom{\dag}}_{\uparrow} & 
		0 & 0 & 0 & 0 & 0 & 0 & 0 & 0 & 0 \\ 
		-t \, {\bf c}^{{\dag}}_{\uparrow} & 	
		0 & 0 & 0 & 0 & 0 & 0 & 0 & 0 & 0 \\ 
		t \, {\bf c}^{\vphantom{\dag}}_{\downarrow} & 
		0 & 0 & 0 & 0 & 0 & 0 & 0 & 0 & 0 \\ 
		-t \, {\bf c}^{{\dag}}_{\downarrow} & 
		0 & 0 & 0 & 0 & 0 & 0 & 0 & 0 & 0 \\  
		\arrayrulecolor{gray} \hline   
		0 & {\bf F} & 0 & 0 & 0 & 0 & 0 & 0 & 0 & 0 \\ 
		0 & 0 & {\bf F} & 0 & 0 & 0 & 0 & 0 & 0 & 0 \\ 
		0 & 0 & 0 & {\bf F} & 0 & 0 & 0 & 0 & 0 & 0 \\ 
		0 & 0 & 0 & 0 & {\bf F} & 0 & 0 & 0 & 0 & 0 \\  
		\arrayrulecolor{gray} \hline   
		\varepsilon_i \, {\bf n} & 
			0 & 0 & 0 & 0 & 
		{\bf c}^{{\dag}}_{\uparrow}{\bf F} & 
		{\bf c}^{\vphantom{\dag}}_{\uparrow}{\bf F} &
		{\bf c}^{{\dag}}_{\downarrow}{\bf F} & 
		{\bf c}^{\vphantom{\dag}}_{\downarrow}{\bf F} & 
		{\bf 1}
	\end{array} 
	}\right] \, ,
	\arrayrulecolor{black} 
	\nonumber
	\\
	\label{eqn:hybridHoppingMPO}
\end{align} 
with ${{\bf n}={\bf n}_{\downarrow}+{\bf n}_{\uparrow}}$.
For the mapping between the MPO chain and the two-dimensional lattice
given in Fig.~\ref{fig:mapping}, an
additional four channels are needed to connect the appropriate
$c^{\vphantom{\dagger}}_{\sigma}$ and $c^\dagger_{\sigma}$ 
operators between next-nearest-neighbor sites.
The transverse hopping is encoded into the dispersion relation and
is thus local in hybrid space; it can then be treated 
just like the on-site repulsion in the real-space MPO.
For arbitrary  cylinder width, 
four channels are needed for each momentum point,
resulting in a total MPO bond dimension of ${2+4\,L_y}$ 
for the combined terms~(\ref{eqn:HybbardTX}) and~(\ref{eqn:HybbardTY}).

\begin{table}[htpb]
	\def\arraystretch{1.35}
	\centering
		\def\a{${\bf c}^\dagger_\uparrow\,$}
		\def\b{${\bf c}^{\vphantom{\dagger}}_\uparrow\,$}
		\def\c{${\bf c}^\dagger_\downarrow\,$}
		\def\d{${\bf c}^{\vphantom{\dagger}}_\downarrow\,$}
	  \def\e{$0$}
	\begin{tabularx}{8.6cm}{
		>{\centering\arraybackslash}p{0.7cm} |
		>{\centering\arraybackslash}p{1.3cm} 
		>{\centering\arraybackslash}p{0.4cm} 
		>{\centering\arraybackslash}p{0.7cm} |
		>{\centering\arraybackslash}p{1.3cm} 
		>{\centering\arraybackslash}p{0.4cm} 
		>{\centering\arraybackslash}p{0.7cm} |
		>{\centering\arraybackslash}p{1.3cm}}
		\hline\hline 
		\multicolumn{2}{c}{\underline{four}} & & 
		\multicolumn{2}{c}{\underline{two}} & & 
		\multicolumn{2}{c}{\underline{one}} \\
		1 & final & & 6  & \a\b & & 12 & \a \\
		\multicolumn{2}{c}{\underline{three}} & & 7  & \a\c & & 13 & \b \\
		2 & \a\b\c & & 8  & \a\d & & 14 & \c \\
		3 & \a\b\d & & 9  & \b\c & & 15 & \d \\		
		4 & \a\c\d & & 10 & \b\d & & 
		\multicolumn{2}{c}{\underline{zero}} \\
		5 & \b\c\d & & 11 & \c\d & & 16 & initial  \\
		\hline   \hline  
	\end{tabularx}
	\caption{ 
		Labels for MPO channels
		grouped according to the number of included operators
		as used for the MPO matrix in Eq.~(\ref{eqn:hybridInteractionMPO}).
	}
	\label{tab:mpochannels}
\end{table}
The last term~(\ref{eqn:HybbardU}) is more complicated to implement,
requiring multiple steps.
For fixed $x$, the sum~(\ref{eqn:HybbardU}) can alternatively be written 
as a sum over four momenta,
\begin{equation}
	\label{eqn:HybbardU2}
	U/L_y\,
	\sum_{k\,k'\,p\,p'} 
	c^\dagger_{k\,\uparrow} \,
	c^{\vphantom{\dagger}}_{k'\,\uparrow} \,
	c^\dagger_{p\,\downarrow} \,  
	c^{\vphantom{\dagger}}_{p'\,\downarrow} \, 
	\delta_{k-k'+p-p'} \, ,
\end{equation}
where the $\delta$-function ensures momentum conservation.
Note that we have dropped the $x$ index
and the second-level $y$  index for compactness.
Furthermore, if we neglect momentum conservation for the moment, 
we obtain a relatively simple sum that contains all possible arrangements
of the four operators
  ${(
	{c^\dagger_\uparrow},
	{c^{\vphantom{\dagger}}_\uparrow},
	{c^\dagger_\downarrow},
  	{c^{\vphantom{\dagger}}_\downarrow})}$
within the sites of one ring of the cylinder:
\begin{equation}
	\label{eqn:HybbardU3}
	U/L_y\,
	\sum_{k\,k'\,p\,p'} 
	c^\dagger_{k\,\uparrow} \,
	c^{\vphantom{\dagger}}_{k'\,\uparrow} \,
	c^\dagger_{p\,\downarrow} \, 
	c^{\vphantom{\dagger}}_{p'\,\downarrow} \, .
\end{equation}
This sum can be represented as an MPO
in a compact form using $16$ channels,
which we label in the following way:
the initial channel designates ``no operator'',
the target channel stands for ``all four operators'',
and the other $14$ channels represent
all possible one-, two-, and three-operator combinations 
(disregarding order and repetition).
The complete set of channels is given in Table~\ref{tab:mpochannels}.
The lower triangular part of the MPO matrices then contains all entries
that logically connect this set of channels in that
the matrix element in row $i$ and column $j$ consists of either the 
operator(s) that must be added to the label of the $i$-th row
to obtain the label of the $j$-th column
or zero if this is not possible.
In addition, the matrices must incorporate the appropriate sign that
takes into account 
the fermionic commutation relations with respect to
the ordering of the operators in Eq.~(\ref{eqn:HybbardU3}) within the
MPO chain.
This can be done by assigning 
a minus sign to all matrix elements
that require an odd number of exchanges of operators
to reorder the combined labels of the row plus the element itself
to match the label of the column.
Finally, 
identity operators and fermionic-sign operators must be placed on the diagonal 
so that operators connect over longer distances.
Following these rules,
the resulting MPO matrix
for one single ring of the cylinder reads
\begin{widetext} 
\begin{align}
	W^{[i]}= \hspace{9cm}\nonumber \\
	\arrayrulecolor{gray}	
	\renewcommand{\arraystretch}{1.3}
	\setlength{\arraycolsep}{0.1cm}
	\def\a{{\bf c}^\dagger_\uparrow}
	\def\b{{\bf c}^{\vphantom{\dagger}}_\uparrow}
	\def\c{{\bf c}^\dagger_\downarrow}
	\def\d{{\bf c}^{\vphantom{\dagger}}_\downarrow}	
	\def\e{0}
	\def\u{{\bf 1}}
	\def\f{{\bf F}}
	\left[
	\scalemath{0.98}
	{
	\begin{array}{c|cccc|cccccc|cccc|c} \arrayrulecolor{gray}
		\u & \e & \e & \e & \e &
		\e & \e & \e & \e & \e & \e &
		\e & \e & \e & \e & \e \\ \hline 
		U/L_y\,\d & \f & \e & \e & \e &
		\e & \e & \e & \e & \e & \e & 
		\e & \e & \e & \e & \e \\
		-U/L_y\,\c & \e & \f & \e & \e &
		\e & \e & \e & \e & \e & \e & 
		\e & \e & \e & \e & \e \\
		U/L_y\,\b & \e & \e & \f & \e &
		\e & \e & \e & \e & \e & \e & 
		\e & \e & \e & \e & \e \\
		-U/L_y\,\a & \e & \e & \e & \f &
		\e & \e & \e & \e & \e & \e &
		\e & \e & \e & \e & \e \\ \hline 
		U/L_y\,\c\d & \c\f & \d\f & \e & \e & 
		\u & \e & \e & \e & \e & \e & 
		\e & \e & \e & \e & \e \\
		-U/L_y\,\b\d &-\b\f & \e & \d\f & \e & 
		\e & \u & \e & \e & \e & \e & 
		\e & \e & \e & \e & \e \\
		U/L_y\,\b\c & \e &-\b\f &-\c\f & \e & 
		\e & \e & \u & \e & \e & \e & 
		\e & \e & \e & \e & \e \\ 
		U/L_y\,\a\d & \a\f & \e & \e & \d\f & 
		\e & \e & \e & \u & \e & \e & 
		\e & \e & \e & \e & \e \\
		-U/L_y\,\a\c & \e & \a\f & \e &-\c\f & 
		\e & \e & \e & \e & \u & \e & 
		\e & \e & \e & \e & \e \\
		U/L_y\,\a\b & \e & \e & \a\f &\b\f & 
		\e & \e & \e & \e & \e & \u & 
		\e & \e & \e & \e & \e \\ \hline 
		{U/L_y\,\b\c\d} & {\b\c\f} & \b\d\f & {\c\d\f} & \e & 
		\b & \c & \d & \e & \e & \e & 
		\f & \e & \e & \e & \e \\
		{-U/L_y\,\a\c\d} &-\a\c\f &{-\a\d\f} & \e & {\c\d\f} &
		-\a& \e & \e & \c & \d & \e & 
		\e & \f & \e & \e & \e \\
		{U/L_y\,\a\b\d} & {\a\b\f} & \e &{-\a\d\f} &-\b\d\f & 
		\e &-\a & \e &-\b & \e & \d & 
		\e & \e & \f & \e & \e \\
		{-U/L_y\,\a\b\c} & \e & {\a\b\f} & \a\c\f & {\b\c\f} & 
		\e & \e &-\a & \e &-\b &-\c & 
		\e & \e & \e & \f & \e \\ \hline 
		U/L_y\,\a\b\c\d & {\a\b\c\f} & {\a\b\d\f} &
		{\a\c\d\f} & {\b\c\d\f} & 
		\a\b & \a\c & \a\d & \b\c & \b\d & \c\d & 
		\a\f & \b\f & \c\f & \d\f & \u
	\end{array}
	}
	\right] \, .
	\label{eqn:hybridInteractionMPO}
	\arrayrulecolor{black}
\end{align}
\end{widetext}
Note that fermionic-sign operators must be applied 
to all elements of the matrix 
which have an odd number of operators to the right within the MPO,
i.e., all elements in columns $2$ to $5$ and $12$ to $15$.
Since the sum~(\ref{eqn:HybbardU}) is local in the $x$-direction,
the MPO for the complete cylinder can be constructed by connecting
$L_x$ MPOs for single rings {\em only} through their initial and final channels.

In order to restore momentum conservation,
we now split each of 
the $14$ one-, two-, and three-operator channels
into $L_y$ separate channels, which correspond to the individual
momentum sectors.
Only the initial and target channels must have zero momentum 
and thus do not need to be split.
The original matrix~(\ref{eqn:hybridInteractionMPO}) must then be transformed 
in the following way:
the corner elements of the matrix
remain unaltered,
the other elements within the first and last row or column
become  ${1{\times}L_y}$ and ${L_y{\times}1}$ sub-matrices, respectively,
and all remaining elements become ${L_y{\times}L_y}$ sub matrices.
Within these sub matrices,
the local operators must be arranged according to their momentum 
and the momenta of the column and row channels, i.e.,
the combined momentum of the row channel and the operator  
must match the momentum of the column channel (modulo system width).
Precisely, if we label the new channels with increasing momenta,
an operator with momentum $k$ is placed on the $k$-th upper 
and ${(L_y-k)}$-th lower diagonal.
For example, 
a creation operator on a site with momentum ${k=2}$ 
within a width-6 cylinder becomes
a ${6{\times}6}$ sub-matrix
\begin{equation}
	\def\arraystretch{1.3}
	\def\d{{\bf c}^{\vphantom{\dagger}}_\uparrow}
	\def\c{{\bf c}^{\vphantom{\dagger}}_\downarrow}
	\def\a{{\bf c}^\dagger_\uparrow}
	\def\b{{\bf c}^\dagger_\downarrow}
	\def\e{0}
	\a \rightarrow 
	\left[
	\begin{matrix}
		0 & 0 & \a & 0 & 0 & 0 \\
		0 & 0 & 0 & \a & 0 & 0 \\
		0 & 0 & 0 & 0 & \a & 0 \\
		0 & 0 & 0 & 0 & 0 & \a \\
		\a & 0 & 0 & 0 & 0 & 0 \\
		0 & \a & 0 & 0 & 0 & 0 
	\end{matrix}
	\right] \, . 
\end{equation} 
The resulting MPO is indeed momentum-conserving, 
and its structure encodes the factorization described
in Sec.~\ref{sec:hybrid-space_mpo},
with the one-, two-, and three-operator channels
corresponding to all possible composed operators
to the left and the right side of each bond.

This completes the steps that are necessary to assemble a complete MPO
for the hybrid-space Hamiltonian.
The MPOs for the different parts of the Hamiltonian can be combined 
by simply using distinct sets of channels for each part,
except for the initial and final channels.
In total, one initial channel, one final channel,
${4\,L_y}$ channels for the longitudinal hopping~(\ref{eqn:HybbardTX}),
and ${14\,L_y}$ channels for the nonlocal 
Hubbard interaction~(\ref{eqn:HybbardU}) are required,
resulting in a total virtual bond dimension of ${2+18\,L_y}$.

The final MPO contains many superfluous nonzero elements,
which should be eliminated to prevent unnecessary calculations.
Since the initial and final channels have zero momentum,
many paths through the MPO described above are dead ends
that cannot contribute to the final results in any meaningful calculation,
i.e., they are either not connected to the initial channel to the left
or have no connection to the target channel to the right.
One can eliminate all elements that are part of such dead ends
and thus significantly reduce the effective average 
virtual bond dimension of the MPO (see Table~\ref{tab:mpo_dimensions}).
Furthermore, it is possible to reuse the one-operator channels
of the interaction MPO to implement the longitudinal hopping
in a more economical way; 
however,
this makes the structure of the MPO more complicated
and results only in minor savings ${(\approx5\%)}$ of
computational time and memory costs.

\FloatBarrier
\section{Complete energy tables}
\label{app:energies}
\FloatBarrier

Tables~\ref{tab:energiesU4} and~\ref{tab:energiesU8} list the
ground-state energies for the two-dimensional Hubbard model at
${n=0.875}$ and ${U/t=4.0}$ and ${U/t=8.0}$, respectively.

\begin{table}[htpb]
	\def\arraystretch{1.25}
	\small
	\centering
	\begin{tabularx}{8.6cm}{
		>{\centering\arraybackslash}p{1.4cm} 
		>{\centering\arraybackslash}p{3.0cm} 
		>{\centering\arraybackslash}p{1.4cm} 
		>{\centering\arraybackslash}p{2.0cm} 
  }
		\hline\hline 
		$L_x{\times}L_y$ & 
		boundary conditions & 
		$\lambda$ & 
		$E_0/t$ \\ \hline 
		{$16{\times}4$}     & PBC & $8.0$ & $-1.018413(2)$  \\
		{$16{\times}4$}     & ABC & $8.0$ & $-1.001587(4)$  \\	\hline 
		{$32{\times}4$}     & PBC & $8.0$ & $-1.028614(6)$  \\
		{$32{\times}4$}     & ABC & $8.0$ & $-1.010893(3)$  \\	\hline 
		{$48{\times}4$}     & PBC & $8.0$ & $-1.032078(7)$  \\
		{$48{\times}4$}     & ABC & $8.0$ & $-1.013997(5)$  \\	\hline 
		{$64{\times}4$}     & PBC & $8.0$ & $-1.033814(4)$   \\ 
		{$64{\times}4$}     & ABC & $8.0$ & $-1.015549(5)$  \\ \hline 
		{$\infty{\times}4$} & PBC & $8.0$ & $-1.03891(3)$   \\ 
		{$\infty{\times}4$} & ABC & $8.0$ & $-1.020202(5)$  \\ \hline 
		{$16{\times}6$}     & PBC & $8.0$ & $-1.0097(5)$   \\
		{$16{\times}6$}     & ABC & $8.0$ & $-1.0088(1)$  \\ \hline 
		{$32{\times}6$}     & PBC & $8.0$ & $-1.0191(2)$   \\
		{$32{\times}6$}     & ABC & $8.0$ & $-1.0184(1)$ \\ \hline 
		{$48{\times}6$}     & PBC & $8.0$ & $-1.0223(2)$   \\
		{$48{\times}6$}     & ABC & $8.0$ & $-1.0214(2)$  \\ \hline 
		{$64{\times}6$}     & PBC & $8.0$ & $-1.0240(3)$   \\
		{$64{\times}6$}     & ABC & $8.0$ & $-1.0230(4)$  \\ \hline 
		{$\infty{\times}6$} & PBC & $8.0$ & $-1.0286(4)$      \\
		{$\infty{\times}6$} & ABC & $8.0$ & $-1.0277(3)$  \\ \hline\hline
	\end{tabularx}
		\caption{ 
			Zero-truncation-error extrapolated ground-state energies 
			of Hubbard cylinders for ${U/t=4.0}$ at ${n=0.875}$ filling
			for different system sizes
			and transverse boundary conditions.
			For width 6, a pinning field 
			was used to stabilize the different stripe configuration
			during the initial DMRG sweeps.
		}
	\label{tab:energiesU4}
\end{table}

\begin{table}[htpb]
	\def\arraystretch{1.25}
	\small
	\centering
	\begin{tabularx}{8.6cm}{
		>{\centering\arraybackslash}p{1.4cm} 
		>{\centering\arraybackslash}p{3.0cm} 
		>{\centering\arraybackslash}p{1.4cm} 
		>{\centering\arraybackslash}p{2.0cm} 
  }
		\hline\hline 
		$L_x{\times}L_y$ & 
		boundary conditions & 
		$\lambda$ & 
		$E_0/t$ \\ \hline 
		{$16{\times}4$}     & PBC & $8.0$ & $-0.75114(2)$  \\
		{$16{\times}4$}     & ABC & $8.0$ & $-0.74712(2)$  \\	\hline 
		{$32{\times}4$}     & PBC & $8.0$  & $-0.75841(2)$  \\
		{$32{\times}4$}     & ABC & $8.0$ & $-0.75382(3)$  \\	\hline 
		{$48{\times}4$}     & PBC & $8.0$ & $-0.76079(2)$  \\
		{$48{\times}4$}     & ABC & $8.0$  & $-0.75604(4)$  \\	\hline 
		{$64{\times}4$}     & PBC & $8.0$  & $-0.7621(5)$   \\ 
		{$64{\times}4$}     & ABC & $8.0$  & $-0.75725(6)$  \\ \hline 
		{$\infty{\times}4$} & PBC & $8.0$ & $-0.7657(3)$   \\ 
		{$\infty{\times}4$} & ABC & $8.0$ & $-0.76057(7)$  \\ \hline 
		{$16{\times}6$}     & PBC & $8.0$ & $-0.7481(2)$   \\
		{$16{\times}6$}     & PBC & $5.3$ & $-0.74745(2)$  \\ \hline 
		{$32{\times}6$}     & PBC & $8.0$            & $-0.7556(7)$   \\
		{$32{\times}6$}     & PBC & $5.3$ & $-0.754702(3)$ \\ \hline 
		{$48{\times}6$}     & PBC & $8.0$            & $-0.7577(3)$   \\
		{$48{\times}6$}     & PBC & $5.3$ & $-0.75727(1)$  \\ \hline 
		{$64{\times}6$}     & PBC & $8.0$            & $-0.7591(2)$   \\
		{$64{\times}6$}     & PBC & $5.3$ & $-0.75842(4)$  \\ \hline 
		{$\infty{\times}6$} & PBC & $8.0$            & $-0.7627(5)$      \\
		{$\infty{\times}6$} & PBC & $5.3$ & $-0.76210(5)$  \\ \hline\hline
	\end{tabularx}
		\caption{ 
			Zero-truncation-error extrapolated ground-state energies 
			of Hubbard cylinders for ${U/t=8.0}$ at ${n=0.875}$ filling
			for different stripe patterns, system sizes, 
			and transverse boundary conditions.
			For width 6, a pinning field 
			was used to stabilize the different stripe configuration;
			the energy contribution of the pinning-field 
			was subtracted afterwards.
		}
	\label{tab:energiesU8}
\end{table}

\FloatBarrier
%

\end{document}